\documentclass[a4paper,fleqn,usenatbib]{mnras}

\usepackage{newtxtext,newtxmath}
\usepackage[T1]{fontenc}
\usepackage{ae,aecompl}

\usepackage{hyperref}
\hypersetup{
    colorlinks,
    citecolor=blue,
    filecolor=black,
    linkcolor=blue,
    urlcolor=black
}
\usepackage{multirow}
\usepackage[dvipsnames]{xcolor} %for colored texts
\usepackage{amssymb} %for math symbols
\usepackage{amsmath} %for math symbols
\usepackage{wasysym} %weird symbols such as astrosun
\usepackage{subfiles} %organize paper into sub folders
\usepackage{subfigure} %to allow sub figures
\usepackage{wrapfig} %to embed a figure within texts
\usepackage{caption} %to have captions i.e. to figures
\usepackage{pdfpages} %to include pdf pictures
\usepackage{textcomp} %to write math symbols as text such as degrees.
\usepackage{diagbox}

\newcommand\fCK{f\textsubscript{\text{CK}}}
\newcommand\fCN{f\textsubscript{\text{CN}}}
\newcommand\Mbh{M\textsubscript{\text{bh}}}
\newcommand\fgout{f\textsubscript{\text{g,out}}}
\newcommand\Rout{R\textsubscript{\text{out}}}
\newcommand\Rin{R\textsubscript{\text{in}}}
\newcommand\Rcric{R\textsubscript{\text{cric}}}
\newcommand\Sigmamp{\Sigma\textsubscript{\text{mp}}}
\newcommand\Sigmastardot{\dot{\Sigma}\textsubscript{*}}
\newcommand\Teff{T\textsubscript{\text{eff}}}
\newcommand\Pmp{P\textsubscript{\text{mp}}}
\newcommand\deltaPE{\delta P\textsubscript{\text{E}}}
\newcommand\deltaPc{\delta P\textsubscript{\text{c}}}
\newcommand\barA{\bar{\text{A}}}

\newcommand\Pgas{P\textsubscript{\text{gas}}}
\newcommand\Prad{P\textsubscript{\text{rad}}}
\newcommand\Pturb{P\textsubscript{\text{turb}}}
\newcommand\vturb{v\textsubscript{\text{turb}}}

\newcommand\tsup[1]{\textsuperscript{#1}}

\newcommand{\percmcube}{\text{cm}\tsup{-3}}
\newcommand{\gpercmsqr}{\,g\,cm$^{-2}$\,}
\newcommand{\percmsqr}{\text{cm}\tsup{-2}}
\newcommand{\cmsqrperg}{\,cm$^{2}$\,g$^{-1}$\,}

\title[Vertical Structure of NSDs]
  {Modeling the Vertical Structure of Nuclear Starburst Discs: A Possible Source of AGN Obscuration at $z\sim 1$}
\author[R.\ Gohil and D.\ R.\ Ballantyne]
  {R.~Gohil\thanks{raj.gohil07@gmail.com} and D.~R.~Ballantyne\\Center
    for Relativistic Astrophysics, School of Physics, Georgia
    Institute of Technology, 837 State Street, Atlanta, GA 30332-0430,
    USA}

\pagerange{\pageref{firstpage}--\pageref{lastpage}}
\pubyear{2017}

\begin{document}
\label{firstpage}
\pagerange{\pageref{firstpage}--\pageref{lastpage}}
\maketitle

\begin{abstract}
Nuclear starburst discs (NSDs) are star-forming discs that may be residing in the nuclear regions of active galaxies at intermediate redshifts. One dimensional (1D) analytical models developed by Thompson et al. (2005) show that these discs can possess an inflationary atmosphere when dust is sublimated on parsec scales. This make NSDs a viable source for AGN obscuration. We model the two dimensional (2D) structure of NSDs using an iterative method in order to compute the explicit vertical solutions for a given annulus. These solutions satisfy energy and hydrostatic balance, as well as the radiative transfer equation. In comparison to the 1D model, the 2D calculation predicts a less extensive expansion of the atmosphere by orders of magnitude at the parsec/sub-parsec scale, but the new scale-height $h$ may still exceed the radial distance $R$ for various physical conditions. A total of 192 NSD models are computed across the input parameter space in order to predict distributions of a line of sight column density $N_H$. Assuming a random distribution of input parameters, the statistics yield 56\% of Type 1, 23\% of Compton-thin Type 2s (CN), and 21\% of Compton-thick (CK) AGNs. Depending on a viewing angle ($\theta$) of a particular NSD (fixed physical conditions), any central AGN can appear to be Type 1, CN, or CK which is consistent with the basic unification theory of AGNs. Our results show that $\log[N_H(\percmsqr)]\in$ [23,25.5] can be oriented at any $\theta$ from 0\textdegree\, to $\approx$80\textdegree\, due to the degeneracy in the input parameters.
%\vspace{1.5cm}
\end{abstract}


\section{Introduction}
Active galactic nuclei (AGNs) are extremely luminous regions, more than a billion times the solar luminosity, in galaxies. They emit energy across a large portion of the electromagnetic spectrum from radio waves to X-rays through accretion onto a supermassive black hole \citep{balbus03}. The majority of AGNs discovered in the thousands by \textit{Chandra} and \textit{XMM-Newton} show the presence of obscuration by dust and gas \citep{brandt05,gilli07,alexander12,lamassa16}. Infrared \citep{thatte97,burtscher13} and X-ray based observations \citep{elvis04,risaliti05} point the location of this obscuration at the parsec scale. The simple unification scheme of AGNs suggests that all the variance in the obscuration are due to the difference in viewing angle ($\theta$) of AGNs with respect to an observer, but they are intrinsically the same objects \citep{antonucci93,netzer15}. The mechanisms and physical conditions responsible behind this obscuration is still not well understood, but they seem to be dependent on AGN luminosity \citep{lawrence10} and, perhaps, redshift \citep{ballantyne06,hasinger08}.

To explain the obscuration at the parsec scale, many possible mechanisms have been proposed. The detection of polarized broad lines in Seyfert 2 galaxies \citep[e.g.,][]{antonucci85} led to the modeling of a simple uniform toroidal absorber \citep{krolik86,krolik88,pier92}. In addition, many groups have proposed a geometrically thick torus supported by infrared (IR) radiation pressure \citep{krolik07,dorodnitsyn11,dorodnitsyn12,chan16,dorodnitsyn16}.  Another possible model is a torus with turbulence pressure due to supernovae or stellar winds \citep{wada05,watabe05}. Moreover, many authors have suggested a warped/tilted-disc to explain a range of obscuration \citep{nayakshin05,caproni06,lawrence10}. In recent years, there has been some work done in modeling a clumpy torus \citep{honig07,nenkova08,honig10} partly due to the observed column density variability seen in X-rays \citep{risaliti02}. Another possible source for the observed obscuration is nuclear starburst regions \citep{fabian98,wada02,thompson05,ballantyne08,hopkins16}. In this paper, we explore the effect of nuclear starburst discs (NSDs) on the AGN obscuration.

More than 50 years has passed since the discovery of the cosmic X-ray background (CXB) \citep{giacconi62} and it is still not entirely resolved into discrete sources. The CXB is the spectral energy distribution (SED) which is characterized by a power law with the photon index of $\Gamma=1.4$ in the 2-10 keV band and a peak in the $10-30$ keV band \citep{gruber99}. The SED up to a few keV energies can be explained by integrating point-like sources \citep{worsley05} and many of these sources are observed to be AGNs \citep[e.g.,][]{mushotzky00,bauer04}. Understanding the distribution of column density along a line of sight $N_H$ is an important aspect in modeling the CXB spectrum \citep{gilli07} since the relative fraction of unobscured (Type 1, $N_H$ $<$ 10\tsup{22} cm\tsup{-2}), Compton-thin Type 2s (CN, 10\tsup{22} cm\tsup{-2} $\leq$ $N_H$ $<$ 10\tsup{24} cm\tsup{-2}), and Compton-thick (CK, $N_H \geq$  10\tsup{24} cm\tsup{-2}) AGNs determine the shape of the spectrum. In particular, the observed peak of the CXB spectrum requires a significant number of CK AGNs at moderate redshifts \citep{comastri95,gilli01,churazov07,moretti09}. Many studies have shown that 10\% to 25\% of CK AGNs are required in order to produce the observed peak of the CXB near 30 keV \citep{draper09,akylas12,ueda14}. However, \cite{akylas12} shows that the fraction of CK AGNs ($\fCK$) can range from 5\% to 50\% due to the degeneracy of the input parameters in the modeling of the CXB spectrum. If NSDs are the potential source for producing the CXB peak, we can place a constraint on $\fCK$ by modeling the 2D structure of NSDs.

Many reasons point toward NSDs as a potential source of obscuration. Young stellar populations are found in the inner regions of nearby AGNs \citep[e.g.,][]{gonzalez01,gu01,cid04,storchi05,ruschel17} including Seyfert galaxies \citep{davies07}. \cite{davies07} highlights a possible causality between star formation (SF) and AGN activity from studies of nine nearby Seyfert galaxies: the AGN activity occurs later in time separated by 50-200 Myrs from the peak of the SF rate. Such a causality suggests a possible fueling of gas toward a central black hole by stellar winds and supernovae \citep[e.g.,][]{vollmer08,hopkins12}, perhaps indicating a strong coupling of SF, stellar winds/supernovae, and the AGN activity. With a simple 1D model of NSDs, \cite{thompson05} shows a region near dust sublimation can inflate to $h\sim R$ which can obscure the incoming AGN irradiation. Such a mechanism can build a bridge among star-forming regions, AGN activity, and the obscuration. These reports provide further motivation to study NSDs in detail. Multi-dimensional modeling of NSD is important in order to resolve the dependency of $N_H$ on the orientation angle $\theta$ of AGNs with respect to an observer.

In order to place a proper constraint on the column density $N_H$ and the fractions of CN Type 2s ($\fCN$) and CK AGNs, the vertical structure of NSDs is required. In the past, the hydrostatic structure of a disc has been computed for accretion discs using an iterative method \citep{hubeny90}. Using similar procedures as described by \cite{hubeny90} and \cite{hubeny98}, we compute 2D structure of NSDs under various physical conditions. Later, we study the dependency of $N_H$ on $\theta$ and also place a constraint on $\fCN$ and $\fCK$ using these models. These are outcomes of the NSD theory which can be used to test whether these discs can plausibly obscure and fuel Seyfert galaxies at $z\sim 1$. (Starburst discs may be a more prominent source for AGN obscuration when a large gas fraction is available in galaxies \citep{ballantyne08}-conditions that are more favorable at intermediate redshifts.)

Sect. 2 reviews the one-dimensional theory of NSDs. Sect. 3 describes the methodology of vertical structure. In Sect. 4, we provide and discuss the results including caveats and future work. Then, the paper is concluded in Sect. 5. Appendix A provides details on the opacity calculation used in the modeling.

\begin{table}
\centering
\caption{Description of symbols which are used frequently in the paper.}
\label{table:symbols}
\begin{tabular}{cc}\hline\hline
Symbol & Description\\\hline\hline
$c$ & Speed of light\\
$c_s$ & Total speed of sound\\
$\fCK$ & Fraction of Compton-thick AGNs\\
$\fCN$ & Fraction of Compton-thin AGNs\\
$\fgout$ & Gas fraction at the outer radius\\
$G$ & Gravitational constant\\
$h$ & Surface scale-height\\
$\kappa$ & Opacity\\
$k_B$ & Boltzmann constant\\
$m$ & Radial Mach number\\
$\Mbh$ & A black hole mass\\
$m_p$ & Mass of proton\\
$N_H$ & Column density along a line of sight\\
$\Omega$ & Keplerian angular frequency of a disc\\
$\Pgas$ &  Gas pressure\\
$\Prad$ & Radiation pressure\\
$\Pturb$ & Turbulence pressure\\
$P$ & Total pressure\\
$Q$ & Toomre parameter\\
$\rho$ & Density\\
$R$ & Radial distance at $z=0$\\
$\Rcric$ & Distance where the largest inflation occurs in a disc\\
$\Rout$ & Size of a disc\\
$\sigma_{\text{sb}}$ & Stefan-Boltzmann constant\\
$\Sigma$ & Vertically mass integrated column density\\
$\Sigmamp$ & Vertically mass integrated total column density to mid-plane\\
$\sigma$ & Velocity dispersion\\
$\Sigmastardot$ & Star-formation rate density\\
$T$ & Temperature\\
$\tau$ & Optical depth\\
$\Teff$ & Effective temperature\\
$\theta$ & Viewing angle\\
$\vturb$ & Turbulence speed\\
$z$ & Vertical height
\end{tabular}
\end{table}

\section{A Brief Overview of Radial Structure}
The one dimensional (radial structure) model of NSDs in the radiation dominated regime has been computed by \cite{thompson05} for ultra luminous infrared galaxies (ULIRGs). These models require four input parameters: size of the disc $\Rout$, the black hole mass $\Mbh$, the gas fraction at the outer radius $\fgout$, and the Mach number $m\equiv v_r/c_s$, where $v_r$ is the radial velocity and $c_s$ is the total speed of sound. (Variables which are frequently used in the paper are listed in Table~\ref{table:symbols}.) The 1D model has a NSD rotating with the Keplerian frequency $\Omega$ around a central black hole with the mass of $\Mbh$. The $\Omega$ includes the potentials of the black hole and the bulge,
\begin{align}
\Omega(R)=\Big(\frac{GM_{\text{bh}}}{R^3}+\frac{2\sigma^2}{R^2}\Big)^{1/2},
\end{align}
where $G$ is the gravitational constant and $R$ is the radial distance. The dispersion $\sigma$ is computed using the $M_{\text{bh}}-\sigma$ relationship \citep{ferrarese00,gebhardt00,tremaine02},
\begin{align}
M_{\text{bh}}=2\times10^8\Bigg(\frac{\sigma}{200 \text{ km s}^{-1}}\Bigg)^4 M_{\astrosun}.
\end{align}
The stability of a disc against gravitational collapse is always governed by Toomre's parameter $Q$. (The parameter $Q$ is maintained to be 1 because there is an observational evidence in infrared ultraluminous galaxies \citep{downes98} and local spiral galaxies \citep{martin01,leroy08,westfall14} including the Milky Way \citep{binney87,rafikov01} for $Q\simeq$1.) A gas fraction is injected at the outer radius of the disc which is driven towards the black hole at the constant Mach number through a presumed global torque (which can be due to a spiral instability or a bar; \cite{goodman03}). The amount of available gas decreases with the radius as gas depletes into the formation of stars in order to maintain the Toomre stability criteria $Q=1$. Then, the accretion rate ($\dot{M}$) and the SF rate ($\dot{M_*}$) are given by
\begin{align}
\dot{M}(R)=\dot{M}_{\text{out}}-\int_{R}^{R_{\text{out}}}2\pi R'\dot{\Sigma}_{*}(R')dR'
\end{align}
and
\begin{align}
\dot{M_*}(R)=\pi R^2\dot{\Sigma}_*(R),
\end{align}
where $\dot{\Sigma}_*$ is the SF rate density. The radiation pressure from this SF provides a vertical support for the dusty atmosphere against gravity. At some inner radius $R_{\text{in}}$, the energy flux production from viscosity takes over the SF energy flux and the $R_{\text{in}}$ is considered as a transition radius between a NSD and an accretion disc (AD).

In this 1D model, the vertical structure is not computed except the scale-height of a photosphere $h_{\text{ph}}$ is approximated by scaling the mid-plane scale-height $h_{\text{mid}}$ with the ratio of $\kappa_{\text{ph}}$ and $\kappa_{\text{mid}}$:
\begin{align}
\label{eqn:happrox}
& h_{\text{ph}} \approx h_{\text{mid}}\Big(\frac{\kappa_{\text{ph}}}{\kappa_{\text{mid}}}\Big),
\end{align}
where $\kappa_{\text{ph}}$ and $\kappa_{\text{mid}}$ are the opacities at the photosphere and mid-plane, respectively. $\kappa_{\text{ph}}$ is computed by assuming that the temperature at the surface is the same as the effective temperature. With this theory, \cite{thompson05} showed that a large expansion of an atmosphere, $h_{\text{ph}}\sim R$, is possible on parsec scales for certain conditions of ULIRGs. Using the same theory, \cite{ballantyne08} reaches similar conclusions for ``Seyfert-like" conditions.

Here, we relax these assumptions and compute the exact vertical structure of NSDs by solving the coupled equations of hydrostatic balance, radiative transfer, and energy balance.

\section{2D Structure}
\begin{figure*}
\includegraphics[width=0.47\textwidth]{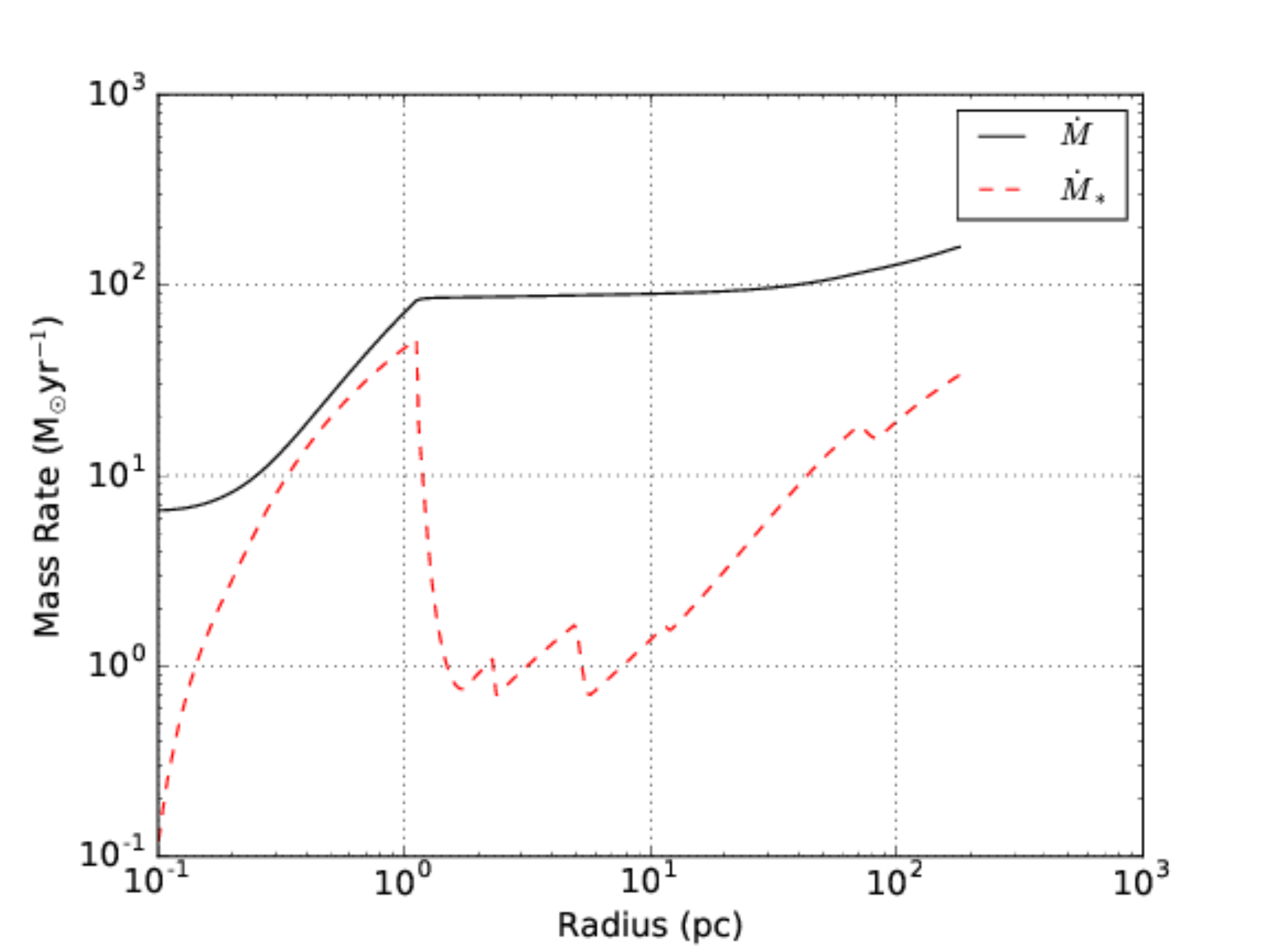}
\includegraphics[width=0.47\textwidth]{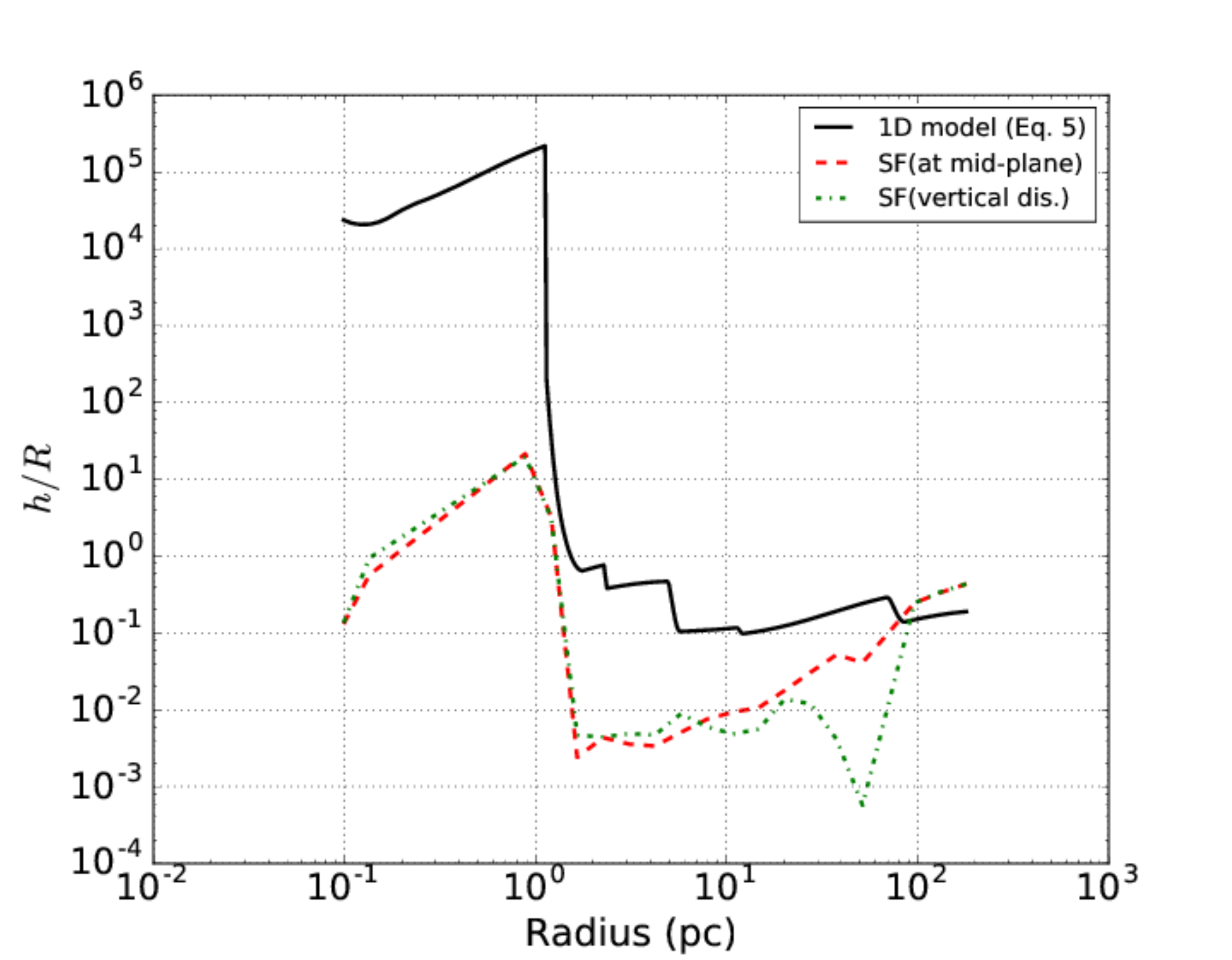}
\centering
\caption{A few important quantities as a function of $R$ for Model A, ($\Mbh$, $\Rout$,$\fgout$, $m$)=(10\tsup{7.5} $M_{\astrosun}$,180.0 pc, 0.8, 0.5). \textit{Left:} The figure shows that the gas is depleted into stars as it accretes toward the black hole in order to achieve the Toomre stability criteria, $Q=1.0$. When the dusty gas reaches $1750K$, graphite grains are destroyed causing a large starburst. This occurs at the critical radius 0.884 pc in order to support the vertical atmosphere by radiation pressure. \textit{Right:} Here, the scale-height $h/R$ computed from the 1D approximation (Eq.~\ref{eqn:happrox}) and the explicit 2D calculation are shown. The dashed-red curve represents the model with SF occurring at the mid-plane and the dashed-dotted-green curve is the model with the vertical distribution of SF (Sect.~\ref{sec:verticaldistribution}). In comparison to an approximated scale-height, the explicit calculation predicts a lower expansion, except at the outer part of the disc. However, near the critical radius, a larger opacity gradient due to the dust sublimation inflates the atmosphere where $h>R$ is achieved.}
\label{fig:modelA_radial}
\end{figure*}

For time-independent conditions, radial and vertical structures are not possible to solve simultaneously due to a lack of prior knowledge on the distribution of any physical quantities. Therefore, the distributions of total energy flux and mass-integrated column density at the mid-plane ($\Sigmamp$) are computed using the \cite{thompson05} model. These distributions are used to compute the vertical structure at every annulus of radial distance. (This means the solutions of each annulus do not communicate with each other.) We assume discs are symmetric around the $z$-axis and $z=0$ plane. NSDs have two energy sources: viscosity and SF. The energy transport is solely by the radiation in the vertical direction. In other words, the conduction and convection heat transports are excluded in modeling. The total radiation flux is parameterized by an effective temperature ($\Teff$) since the energy is transported losslessly to the surface and then radiated away.
\subsection{Vertical Structure of an Annulus}
One can ask what is the vertical structure for a given column of gas with total energy flux at a radial distance $R$ away from a central black hole of mass $\Mbh$. In order to obtain the hydrostatic solution for such a column of gas, the coupled one dimensional time independent radiative transfer equation (RTE), hydrostatic equation (HSE), and energy balance equation (EBE) must be solved.

The RTE governs the distribution of the radiation field within a material which is illustrated in Eq.~\ref{eqn:rte} for a 1D slab:
\begin{align}
\label{eqn:rte}
& \mu\frac{\delta I_\nu(z,\mu)}{\delta z}=\eta_\nu(z,\mu)-\chi_\nu(z,\mu) I_\nu(z,\mu).
\end{align}
Here, $\nu$ is the frequency of light, $\mu$ is the cosine angle, $\eta_\nu$ is the emissivity, and $\chi_\nu$ is the extinction coefficient. To solve the RTE, we make the following approximations:
\begin{enumerate}
\item [1.] a homogeneous plane-parallel geometry (azimuthal symmetry) where properties of a material
only depends on the $z$-direction, i.e. $\chi_\nu$ is isotropic,
\item [2.] isotropic absorption and emission where a source function $S_\nu$ is independent of $\mu$, $S_\nu(\mu,z)=S_\nu(z)$,
\item [3.] local thermodynamical equilibrium (LTE),
\item [4.] grey approximation (frequency integrated radiation field and opacity), and
\item [5.] the Eddington approximations: $f_k(\tau)\equiv\frac{K(\tau)}{J(\tau)}$\ and \ $f_H(\tau=0)\equiv\frac{H(\tau=0)}{J(\tau=0)}$ where $J$, $H$, and $K$ are the first, second, and third angular moments of a radiation field.
\end{enumerate}
Then, the temperature profile computed from the RTE for an energy source is given by
\begin{align}
\label{eqn:tempsolution}
T^4(\tau)=\frac{3}{4}T^4_{\text{eff}}\Big[\tau+q(\tau)\Big]
\end{align}
where $q(\tau)$ is the Hopf function\footnote{The form of $q(\tau)$ can be obtained from Eq. 3.8 in the paper by \cite{hubeny90}.} and $\Teff$ is the total effective temperature. The optical depth $\tau$ is defined as\\

\begin{align}
\tau (\Sigma)& \equiv \int_{\Sigma_1}^{\Sigma}\kappa(\Sigma')d\Sigma'
\end{align}
where $\kappa$ is the opacity and $\Sigma_1$ is the mass-integrated column density of the surface layer. For numerical convenience, $\Sigma$ (which increases monotonically from the surface layer of a slab to its mid-plane) is chosen as the independent variable for all the equations. Then, the vertical distance of a layer from the mid-plane, scale-height $z$, is related to $\Sigma$ with the following equation:
\begin{align}
z(\Sigma)=\int_{\Sigmamp}^{\Sigma} -\frac{1}{\rho(\Sigma')}d\Sigma',
\end{align}
where $\rho$ is the density. Since a NSD has two energy sources, an overall temperature can be computed by adding the fluxes of all the sources, $F_{\text{tot}}=F_{*}+F_{\text{vis}}$, which results in
\begin{align}
T=(T_*^4+T_{\text{vis}}^4)^{1/4}.
\end{align}
Here, $T_*$ and $T_{\text{vis}}$ are temperatures corresponding to the flux produced by viscosity and SF, respectively. Under LTE, $\kappa$ depends on $\rho$, $T$,
\begin{align}
\kappa=\kappa(\rho,T)
\end{align}
and the composition of the slab.  The calculation of $\kappa$ is described in Appendix~\ref{appendix:rmo}.

For the condition of hydrostatic balance, the second order differential form of the HSE,
\begin{align}
\label{eqn:hse}
& \frac{d^2P(\Sigma)}{d\Sigma^2}=-\frac{c_s^2(\Sigma)\Omega^2}{P(\Sigma)}-4\pi G,
\end{align}
is used with the following boundary conditions\footnote{Subscript `1' and `mp' represent
quantities at the surface and mid-plane, respectively.}:
\begin{align*}
%\intertext{with following boundary conditions}\\
\Pmp(\Sigmamp)=\Pmp\ \ \text{and} \ \ P_{1}(\Sigma_{1})=P_{1}.
\end{align*}
Here, $P$ is the total pressure and $\Omega$ is the Keplerian frequency. $P_1$ is computed assuming a constant temperature for $\Sigma<\Sigma_1$ \citep[][section IV(b)]{hubeny90}. $c_s$ is the total speed of sound defined as:
\begin{align}
\label{eqn:CSE}
c_s^2(\Sigma)=\frac{P(\Sigma)}{\rho(\Sigma)}.
\end{align}
The HSE also includes an approximated self-gravity term \citep{paczynski78} and it is solved in the logarithmic space in order to resolve the structure accurately near the surface. The solution of this boundary value problem can be achieved using the finite-difference method.

Finally, the EBE, Eq.~\ref{eqn:ebe}, balances the total pressure with the gas pressure $\Pgas$, radiation pressure $\Prad$, and the turbulence pressure $\Pturb$:
\begin{align}
\label{eqn:ebe}
P(\Sigma)=\Pgas(\Sigma)+\Prad(\Sigma)+\Pturb(\Sigma).
\end{align}
The gas and radiation pressures are computed as
\begin{align}
\Pgas(\Sigma) =\frac{\rho(\Sigma) k_BT(\Sigma)}{m_p}
\end{align}
and
\begin{align}
\Prad(\Sigma) = \frac{4\sigma_{\text{sb}}}{3c}T^4(\Sigma),
\end{align}
while $\Pturb$ is approximated \citep{thompson05} as
\begin{align}
\label{eqn:pturb}
\Pturb(\Sigma) & =1.5\times10^8\Big[\frac{E_0}{10^{51}}\Big]^{13/14}\Big[\frac{\rho(\Sigma)}{m_p}\Big]^{-1/4}\dot{\Sigma}_{*}(\Sigma)\\
            & =\frac{1}{2}\rho(\Sigma) \vturb^2,
\end{align}
where $E_0=$ 10\tsup{51} ergs is the energy released from a supernova. The turbulence speed $\vturb$ is approximated to be a constant in the vertical direction using the density computed from the radial structure if $\Sigmastardot$ is only allowed to occur at the mid-plane. $k_B$, $m_p$, $\sigma_{\text{sb}}$, and $c$ are the Boltzmann constant, mass of the proton, Stefan-Boltzmann constant, and the speed of light, respectively. We also require a column of gas to satisfy the Toomre stability criteria at the mid-plane which is parameterized by $Q$:
\begin{align}
Q(\Sigmamp)=\frac{c_s(\Sigmamp)\Omega}{2\pi G\Sigmamp}=1.0.
\end{align}
Since the only known information about these coupled equations are the boundary conditions, an iterative method is used to solve them.
\subsection{Iterative Method}
To solve the coupled equations described in the previous section, an iterative algorithm is developed which is non-trivial due to the difficulties in achieving convergence. However, the major iterative steps for constructing the 2D structure of a NSD are as follow:
\begin{enumerate}
\item [1.] Using the \cite{thompson05} model, compute the radial distribution of mass-integrated column density at the mid-plane $\Sigma_{\text{mp}}(R)$ and the effective temperature $\Teff(R)$.
\item [2.] Divide the radial distance (from $\Rin$ to $\Rout$) into 50 annuli which are spaced logarithmically.
\item [3.] For a given annulus, we set 300 vertical logarithmic grid points in terms of $\Sigma$ from the surface layer $\Sigma_1=$ 10\tsup{-3}\gpercmsqr to the mid-plane layer $\Sigmamp$.
\item [4.] Compute the initial solution (isothermal profiles) for the energy source embedded only at the mid-plane using a similar procedure to the one described in section IV(a) of \cite{hubeny90}.
\item [4.1.] \label{im:ig_firstlayer} Assume $\kappa(\Sigma_{1})=1.0$ \cmsqrperg and $\rho_{\text{mp}}$ from the radial density profile. Using the analytical isothermal solution, compute $z_1$, $\tau_1$, and $T_1$ in subsequent order. Compute $\kappa_1$ for a given $\rho_1$ and $T_1$ using the procedure described in Appen.~\ref{appendix:rmo}; thereafter, compute the remaining physical quantities (i.e., pressures and speed of sound).
\item [4.2.] Repeat step 4.1 for the rest of the layers to complete the initial vertical profiles. If there is any numerical issue in computing the initial solution, it can be fixed by changing $\rho_{\text{mp}}$ or $\Sigma_1$. For example, the $\Sigma_1$ may be adjusted for a low density environment.
\item [5.] Update the pressure by solving Eq.~\ref{eqn:hse}, where $\Pmp$, $P_1$, and $c_s$ are used from a previous iteration.
\item [6.] Using the updated pressure (or the speed of sound), re-compute $\rho(\Sigma)$, $\kappa(\Sigma)$, $\tau(\Sigma)$, and $T(\Sigma)$ in subsequent order.
\item [7.] Since, the RTE and HSE are not solved simultaneously, the pressure computed from the HSE does not necessarily match with the EBE which induces a discrepancy in pressure, $\deltaPE$. If Eq.~\ref{eqn:CSE} is not consistent with the HSE, it causes another discrepancy in pressure, $\deltaPc$. To reduce $\deltaPE$ and $\deltaPc$, iterate the previous solution by updating $c_s$ using the induced $\deltaPE$ as follows,
\begin{align}
\label{eqn:ucs}
c_s\rightarrow c_s+\delta c_s\ \ \text{where}\ \ \delta c_s=\frac{1}{2}\frac{\deltaPE}{c_s\rho}.
\end{align}
Once $\deltaPE$ and $\deltaPc$ reach their minimum values, move to the next step.
\item [8.] If $\deltaPE$ and $\deltaPc$ are within the set discrepancy of 0.05, skip this step. Otherwise, go back to the step 5.
\item [9.] So far the achieved solution does not necessary satisfy the Toomre criteria ($Q=1$). To maintain this condition at the mid-plane, repeat steps 4-8, but change $\Pmp$ while solving the HSE. $\Pmp$ corresponding to $Q=$1 can be found using the Newton-Raphson method\footnote{$\Pmp=0.9\Prad$ from the initial solution is a good initial guess for the Newton-Raphson method where $\Prad$ is the radiation pressure at the mid-plane.}. Finally, the self-consistent profiles are computed for a given annulus.
\item [10.] Repeat steps 1 to 9 for all the radii of a disc to construct the 2D profiles where $\theta$ is the viewing angle measured from the $z=0$ plane.
\end{enumerate}

\begin{table}
\centering
\caption{We list all the bins in the domain of each input parameter ($\Mbh$, $\Rout$, $\fgout$, $m$). After taking all the possible combination of these bins, a total of 192 models are computed. The Mach number $m$ is chosen such that gas can accrete to the parsec scale and the other parameters represent a possible range of physical conditions for NSDs.}
\label{table:ips}
\begin{tabular}{c c c c c}
log($\Mbh/M_{\astrosun}$) & $\Rout$(pc) & $\fgout$ & $m$ & \\\hline
6.5      & 240  & 0.2   & 0.1  &  \\
7.0      & 180  & 0.4   & 0.3  &  \\
7.5      & 120  & 0.6   & 0.5  &  \\
8.0      & 60   & 0.8   &  --  &
\end{tabular}
\end{table}

Using this developed numerical scheme, we explore NSDs under various conditions (input parameter space) that are shown in Table~\ref{table:ips}. The Mach number $m$ is chosen on the order of a tenth because the gas is not accreted to a parsec scale for a very low $m$ since all the gas is converted into stars in the outer region of the disc and the high Mach number (close to unity or higher) requires an inclusion of additional physics (i.e., shock waves). The size of the disc, the gas fraction at the outer radius, and the mass of the black hole are chosen such that they cover a reasonable physical range of NSDs. Taking all the possible combination of these four input parameters, we compute in total 192 models.

\section{Results \& Discussion}
\begin{figure}
\includegraphics[width=0.45\textwidth]{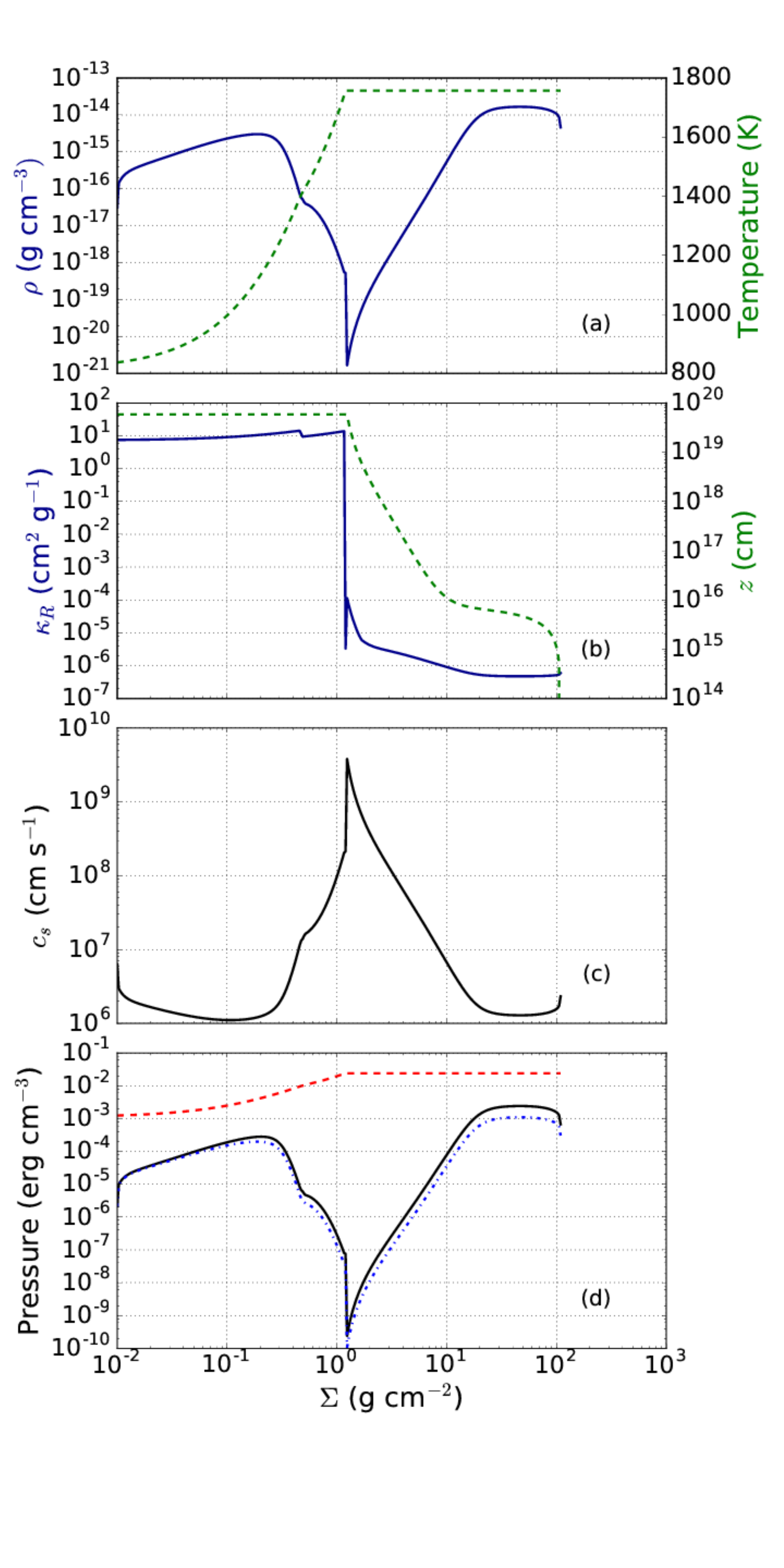}
\centering
\caption{A few important vertical profiles as a function of $\Sigma$ at $\Rcric=0.884$ pc for Model A, ($\Mbh$,$\Rout$,$\fgout$, $m$)=(10\tsup{7.5} $M_{\astrosun}$,180.0 pc, 0.8, 0.5). $\rho$ (solid-blue) and $T$ (dashed-green) increase with $\Sigma$ to achieve the hydrostatic balance as shown in panel (a). Panel (b) shows that when grains sublimate near $\Sigma \approx 1.0$\gpercmsqr, the dust opacity is lost (solid-blue) causing a drop in $\rho$ (solid-blue line in panel (a)).  This increases $c_s$ as shown in the panel (c) which results in the inflation of the surface scale-height (dashed-green line in panel (b)). The solid-black, dashed-red, and dashed-dotted-blue curves in the bottom panel represent $\Pgas$, $\Prad$ and $\Pturb$, respectively. The panel (d) shows that the atmosphere is radiation pressure dominated. A density inversion is developed at the surface; however, it has a negligible effect on the overall $h$ since the inverted column of gas contains less than 1\% of the total column density $\Sigmamp$. A very small density inversion near the mid-plane (panel a) is introduced due the boundary condition of NSDs (input of large energy flux at the mid-plane) and this, in return, causes the inversion of $\Pgas$ which is shown in the panel (d).}
\label{fig:modelA_vertical}
\end{figure}
Out of 192 models, 99 discs show a large expansion of an atmosphere (comparable to $R$ or even higher) on parsec/sub-parsec scale. Here, we present the explicit profiles for one of these discs (hereafter Model A) as an example and its vertical profiles at the critical radius $\Rcric$ (where the largest expansion occurs). Later, we compute a distribution of $N_H$ based on a random selection from sets of $\theta$ and 192 models.

\subsection{Model A}
Model A is the nuclear starburst disc whose $\fgout$, $\Rout$, $m$, and $\Mbh$ are set to 80\%, 180 pc, 0.5, and $10^{7.5} M_{\astrosun}$, respectively.  As gas accretes toward the black hole, the gas is depleted into star-formation in order to maintain the stability criteria (Fig.~\ref{fig:modelA_radial} (left)). In addition, each column of gas becomes denser as gas gets closer to the black hole due to an increase in the vertical gravitational component. This increase in $\Sigma$ enhances the optical depth. As the gas becomes more optically thick, its temperature rises since $T$ is directly related to $\tau$ (which is the consequence of the RTE solution). When the gas crosses the dust sublimation temperature which are 1400K and 1750K for silicate and graphite grains, respectively \citep{kishimoto13}, starbursts occur near $\sim$4.0 pc and $\sim$0.884 pc as shown in Fig.~\ref{fig:modelA_radial} (left). The loss of the dust opacity induces a large opacity gradient. This opacity gradient causes a large expansion of the atmosphere which is illustrated in the right panel of Fig.~\ref{fig:modelA_radial}. This atmosphere is supported by radiation pressure from starbursts (Fig.~\ref{fig:modelA_radial} (left)). The right panel of Fig.~\ref{fig:modelA_radial} compares the approximated surface scale-height (Eq.~\ref{eqn:happrox}) from the \cite{thompson05} model to the explicit calculation. The dashed-red curve assumes that SF occurs at the mid-plane, while the dashed-dotted-green curve represents the model with a vertical distribution of SF (which is discussed further in detail in Sect.~\ref{sec:verticaldistribution}). The exact calculation predicts a lower scale-height ($\approx$ 4 orders of magnitude lower at $\Rcric=0.884$ pc) than the simple approximation except at the outer part of the disc. This discrepancy is due to the inaccuracy in assumptions made under Eq.~\ref{eqn:happrox}: (1) the surface temperature is the same as $\Teff$, (2) a constant $\kappa_{\text{mid}}$ from $z=0$ to $h_{\text{mid}}$, and (3) $\kappa$ increases linearly with $z$. The expansion of the atmosphere at $\Rcric$ is larger than the radial distance by a factor of 21.5 and covers 97\% of the sky (observed from the central black hole).

Fig.~\ref{fig:modelA_vertical} shows several important vertical profiles of a slab at $\Rcric=0.884$ pc for model A. As $\Sigma$ increases from the surface layer ($\Sigma=$10\tsup{-2}\gpercmsqr) to the mid-plane ($\Sigmamp$), density and temperature also increase to achieve hydrostatic balance except when dust sublimates near $\Sigma=1.0$\gpercmsqr which is shown in Fig.~\ref{fig:modelA_vertical} (a). The sublimation of grains results in the immediate drop in the opacity (solid-blue line in Fig.~\ref{fig:modelA_vertical} (b)). This opacity gradient increases the total speed of sound (Fig.~\ref{fig:modelA_vertical} (c)) which, in turn, expands the dusty atmosphere (dashed-green line in Fig.~\ref{fig:modelA_vertical} (b)). Panel (d) shows that radiation pressure (dashed-red curve) dominates throughout the slab over the gas pressure (solid-black curve) and the turbulence pressure (dashed-dotted-blue curve). In such slabs, the density inversion is expected locally since grains are sublimated; however, its effect on the overall scale-height $h$ is insignificant since only a small fraction ($<1\%$) of $\Sigmamp$ is inverted. In addition, the gas pressure curve in panel (d) shows a slight inversion near the mid-plane. This is due to another density inversion caused by the boundary condition of our systems (outburst of energy flux at the mid-plane).

\begin{figure}
\includegraphics[width=0.45\textwidth]{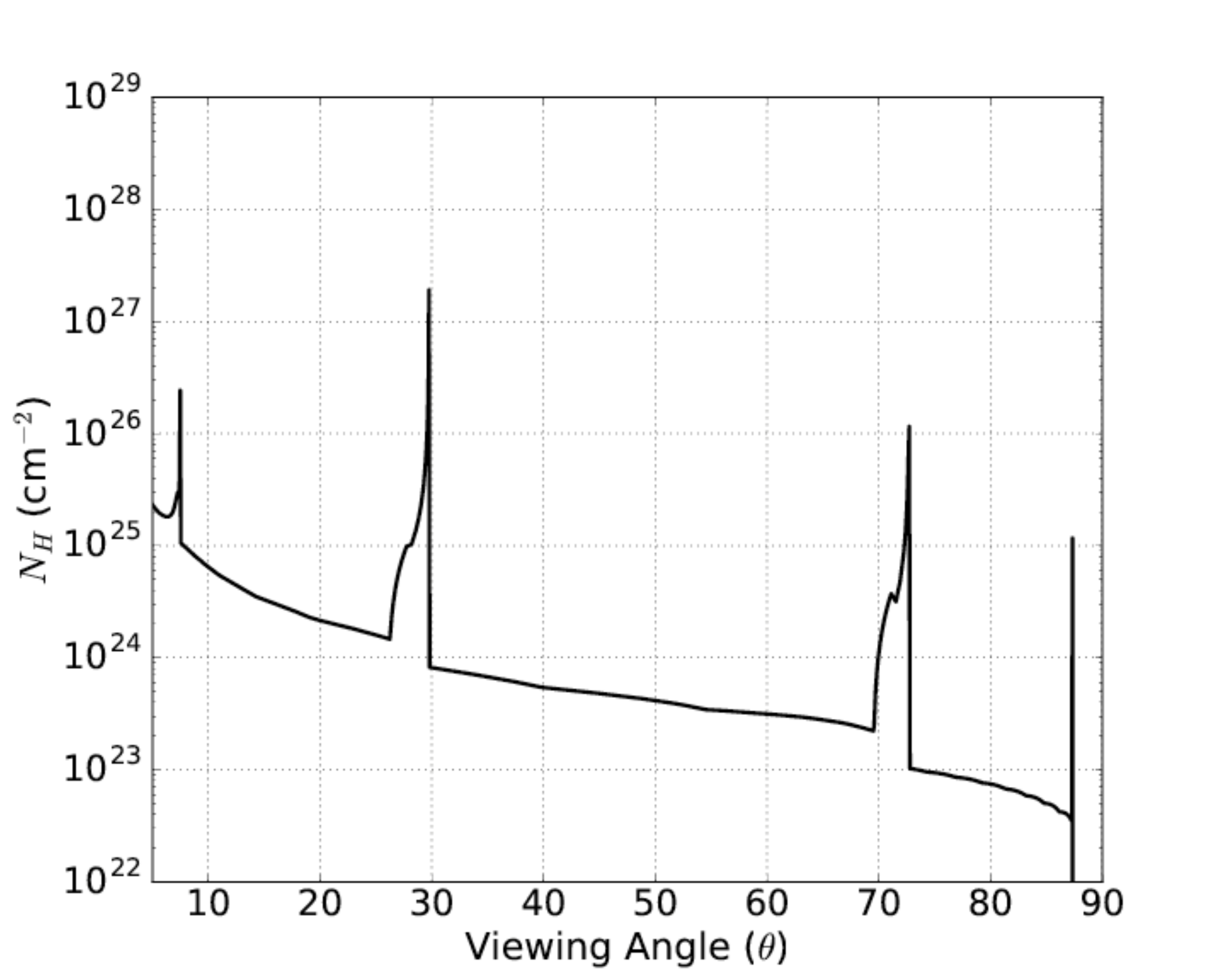}
\centering
\caption{The dependence of $N_H$ as a function of viewing angle $\theta$ for Model A whose radial and vertical profiles are shown in Fig.~\ref{fig:modelA_radial} and Fig.~\ref{fig:modelA_vertical}, respectively. Any central AGN would appear to be CN or CK based on the orientation of the AGN-NSD system with respect to an observer. The three peaks in $N_H$ are due to the density inversion phenomenon which are observed in slabs at $\Rcric$ and beyond.}
\label{fig:NH_theta}
\end{figure}

\begin{figure}
\includegraphics[width=0.46\textwidth]{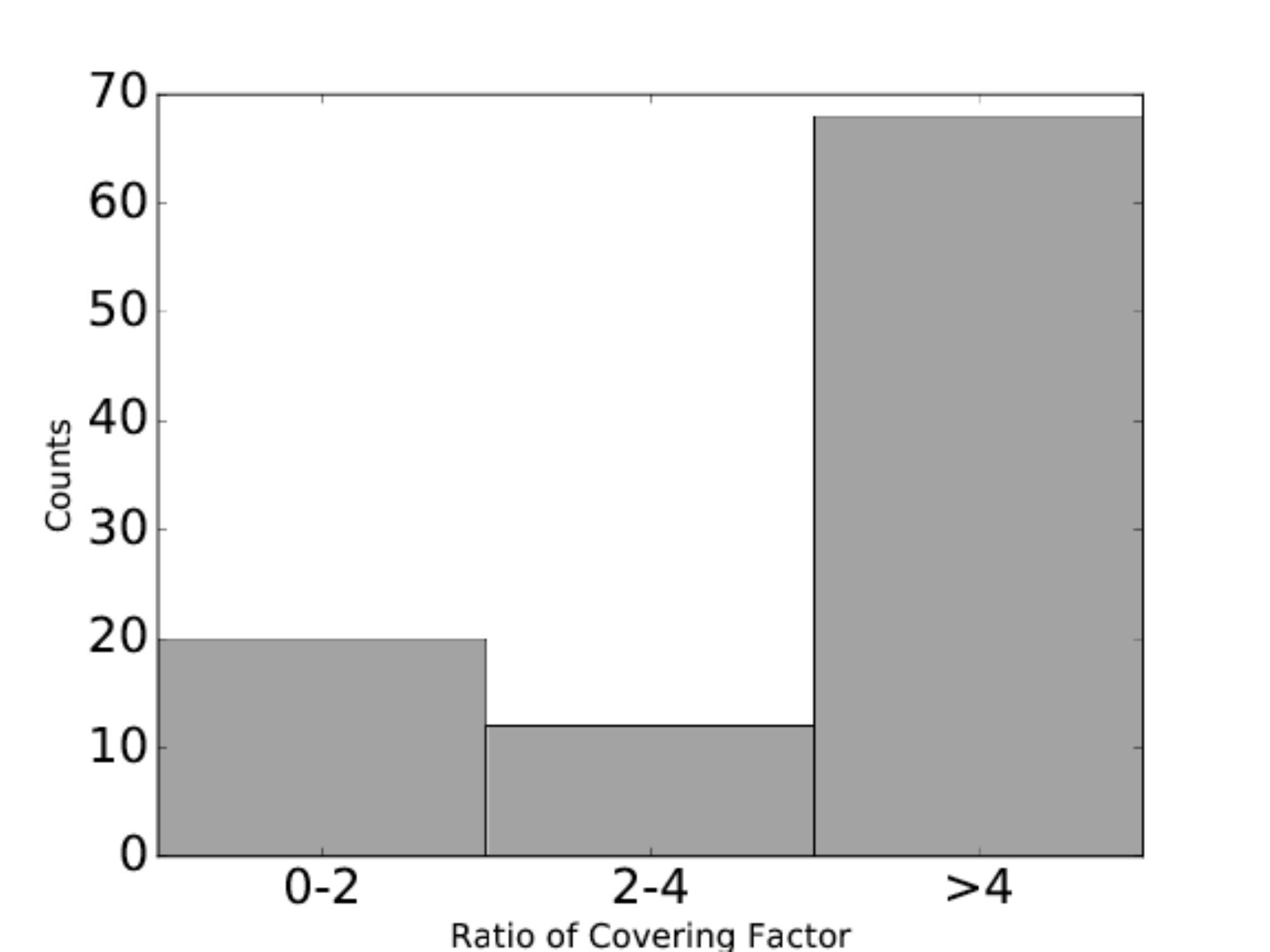}
\centering
\caption{The distribution of the ratio of sky covered by $N_H > 10^{22}$ cm$^{-2}$ gas to that covered by $N_H$ < 10$^{22}$ cm$^{-2}$ gas (i.e., the potential AGN Type 2/Type 1 ratio) for the 99 NSD models that produced a parsec-scale starburst. About 31\% of models (which can vary across the input parameter space) predict covering factors in agreement with the observed ratio. Depending on individual conditions, the fraction of sky covered by a NSD can vary greatly from galaxy to galaxy.}
\label{fig:hist_T21}
\end{figure}

Once the 2D profiles of a disc is calculated successfully, a subsequent task is to compute the column density along a line of sight $N_H$ which is given by
\begin{align}
& N_H \approx \int_{R_{\text{in}}}^{R_{\text{out}}}\rho(\theta,R)\cos^{-1}(\theta) dR/m_p.
\end{align}
For model A, the distribution of $N_H$ as a function of $\theta$ is shown in Fig.~\ref{fig:NH_theta} and shows that $N_H$ decreases with $\theta$. The peaks in $N_H$ are due to the density inversion phenomenon which occurs in slabs where dust is sublimated (Fig.~\ref{fig:modelA_vertical}a). The relative size of these peaks trace the offset of the large opacity gradient among these slabs. The figure illustrates that $N_H$ appears to be CK for $\theta\la$30\textdegree and also near $\theta$ where the peaks occur ($\approx$72\textdegree and 87\textdegree). For any other $\theta$ between $\approx$ 30\textdegree and $\theta_{\text{max}}=87.3$, a central AGN is obscured by the CN medium. For $\theta$>87.3\textdegree, the central AGN appears to be Type 1.
A NSD therefore is consistent with the simple AGN unification model where the line-of-sight into the nucleus determines the observed obscuration properties. Although Model A would imply a Type 2/Type 1 ratio of $\sim$7:1, far higher than the observed ratio $\sim$4:1 \citep{gilli07}), depending on the physical conditions $\theta_{\mathrm{max}}$ can be as low as 40\textdegree, giving a Type 2 to Type 1 ratio of 0.7. The distribution of the obscured to unobscured sky covering factors of all 99 NSD models with an inflationary atmosphere at parsec scale is shown in Fig.~\ref{fig:hist_T21}. The plot shows that, while large ratios are more common in our parameter space, Type 2/Type 1 ratios in line with the observations are found in 31\% of the models. These are the models whose $h/R$ at the critical radius is less than 2 and their physical conditions are scattered across the input parameter space. Therefore, the fraction of sky obscured by the NSD can vary significantly from galaxy to galaxy.

\begin{figure}
\includegraphics[width=0.46\textwidth]{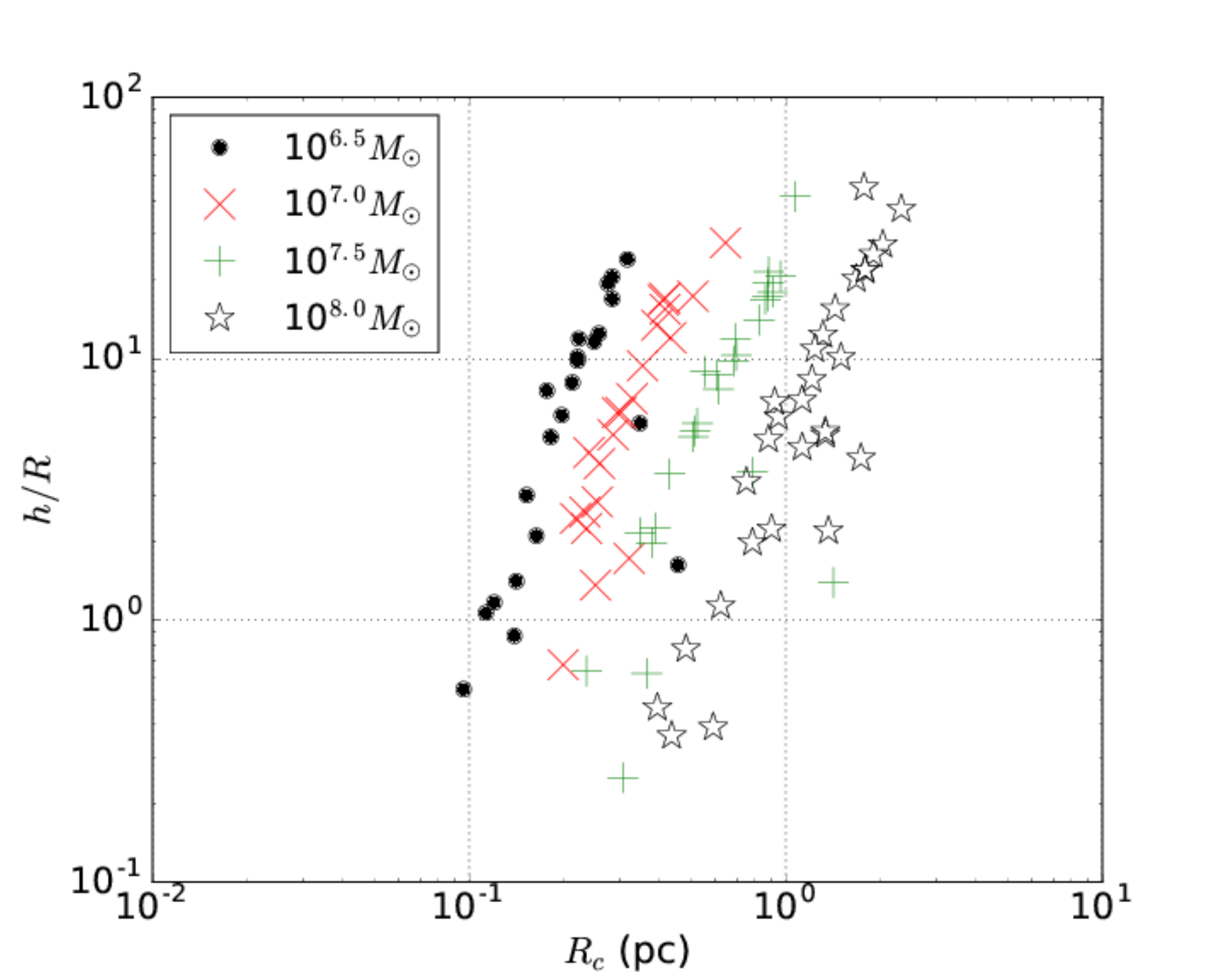}
\centering
\caption{The figure shows $h/R$ inflated due to the dust sublimation at critical radii dependent on $\Mbh$. For a higher $\Mbh$, an expansion occurs earlier due to a relatively stronger vertical gravitational component. This is also the reason why $h/R$ decreases closer to the black hole for a fixed $\Mbh$.}
\label{fig:height}
\end{figure}

\begin{table}
\centering
\caption{A total 99 models out of 192 show the starburst phenomenon causing the large atmosphere expansion at the parsec scale. Here, we show a distribution of these models across the input parameter space. This shows that a smaller disc size, a larger gas fraction and Mach number are in favor of a burst of star-formation.}
\label{table:models_dist}
\begin{tabular}{ccccccc}\hline\hline
&log($\Mbh/M_{\astrosun}$)& 6.5 & 7.0 & 7.5 & 8.0&Total (99)\\\hline\hline
$\Rout$(pc)
&60  & 7 & 6 & 9 & 10 & 32\\
&120 & 5 & 6 & 6 & 8 & 25\\
&180 & 5 & 4 & 6 & 6 & 21\\
&240 & 5 & 5 & 5 & 6 & 21\\\hline
$\fgout$
&0.2 & 0 & 0 & 1 & 3 & 4\\
&0.4 & 5 & 6 & 7 & 8 & 26\\
&0.6 & 8 & 8 & 9 & 9 & 34\\
&0.8 & 9 & 7 & 9 & 10 & 35\\\hline
$m$
&0.1 & 1 & 1 & 2 & 3 & 7\\
&0.3 & 9 & 10 & 11& 13 & 43 \\
&0.5 & 12 & 10 & 13 & 14 & 49\\
\hline\hline
%& &  &  &  &  & Total (99)\\\hline\hline
\end{tabular}
\end{table}

\subsection{All Models: scale-height $h$}
A total of 99 out of 192 models (52\%) show the inflation of an atmosphere at a critical radius which is illustrated in Fig.~\ref{fig:height}. It shows the dependency of $h/R$ on the black hole mass and the critical radius. The black, red, green and white data points at the parsec scale show that the critical radius gets closer to a black hole as $\Mbh$ decreases. This is expected since the gas has to accrete more inward in order to reach to the dust sublimation temperature for the lower $\Mbh$. Depending on the physical conditions, $h/R$ at the sub-parsec scale can range on the order of few tenths to tens. These large expanded vertical structure can obscure the incoming AGN light. Especially, since these atmospheres are dusty, the UV/optical photons will be reprocessed into IR photons. For a given black hole mass, the $\Rcric$ increases as a function of $\fgout$ since the optical depth reaches the sublimation temperature quicker due to the availability of gas. The amount of expansion at these critical radii are mainly governed by the magnitude of starbursts occurring at the mid-plane. Table ~\ref{table:models_dist} shows the distribution of 99 models across the input parameter space. In order to acquire an expanded atmosphere at the parsec scale, a smaller size disc, larger gas fraction, and higher Mach number are favorable conditions which is in the agreement with the work of \cite{ballantyne08}.

\begin{figure}
\includegraphics[width=0.46\textwidth]{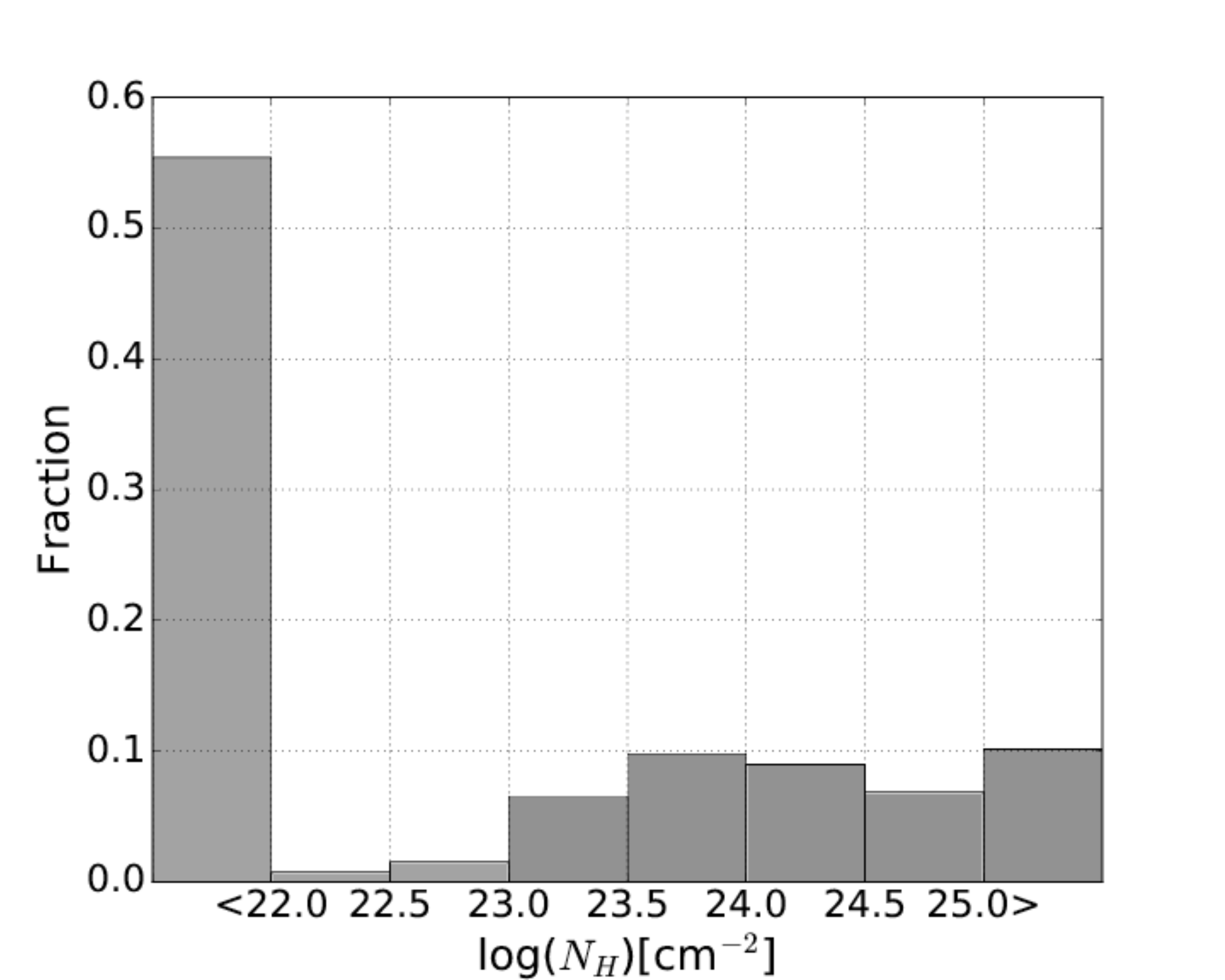}
\includegraphics[width=0.45\textwidth]{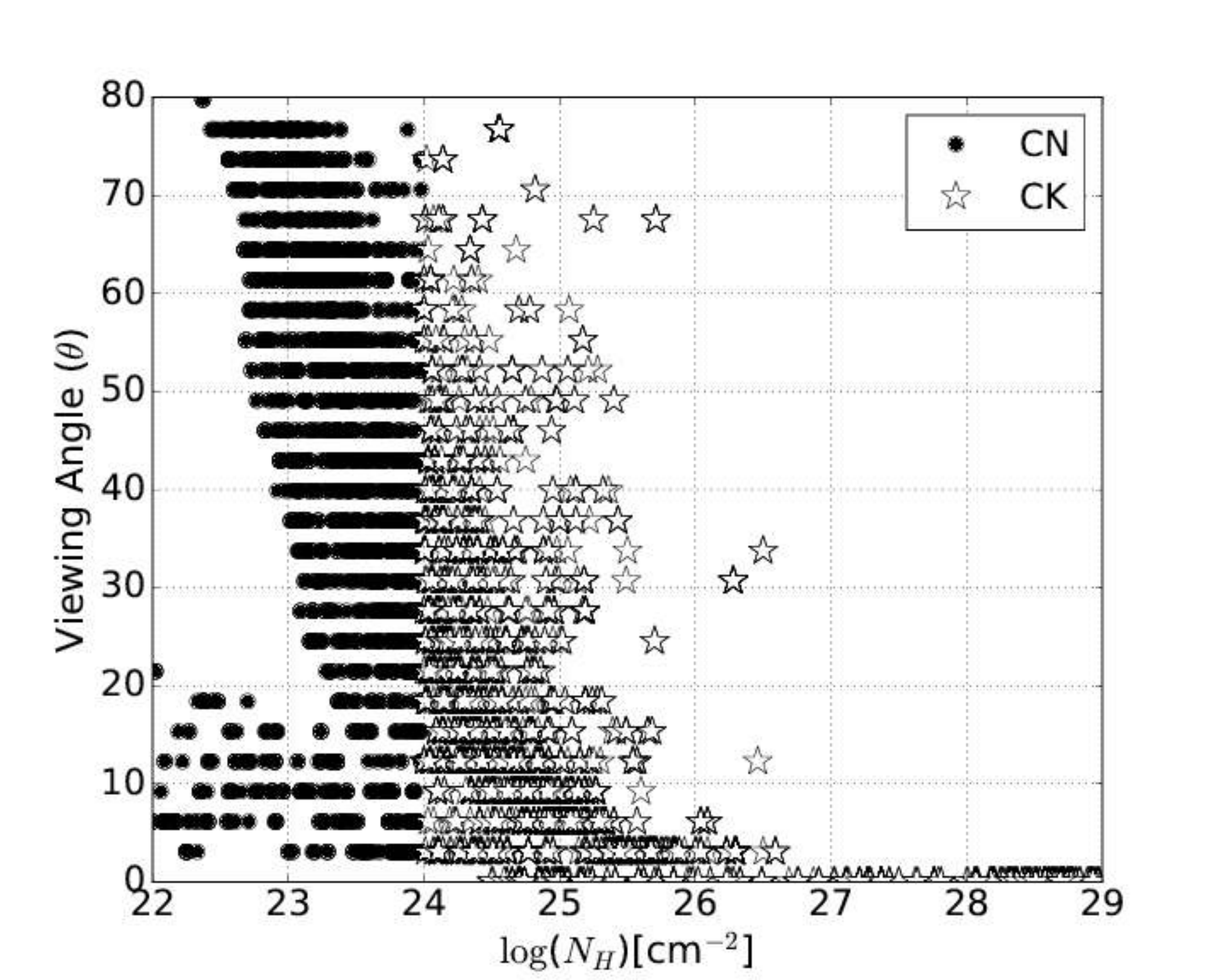}
\centering
\caption{\textit{Top:} A histogram of the AGN fraction as a function of $N_H$ based on a random selection. The fraction of Type 1, CN Type 2s, and CK AGNs are 0.56, 0.23, and 0.21, respectively. The distribution of heavily obscured AGNs peaks near $\log[N_H(\percmsqr)]=23.5$. \textit{Bottom:} Scatter plot of CN and CK AGNs illustrating the dependency of $N_H$ on the orientation of NSD-AGN systems based on the random selection. Due to the degeneracy in the input parameter space, the NSD-AGN systems for a given $N_H$ spans almost an entire domain of $\theta$ for CK and heavily obscured CN AGNs. The gas dominated models do not show the large expansion ($\theta\la 20$\textdegree) which are presented by data points near $\log[N_H(\percmsqr)]$=22.5.}
\label{fig:stat}
\end{figure}

\subsection{$N_H$ Distribution}
To estimate the distribution of the column density observed along a line of sight ($N_H$) from a large sample of NSDs, the domain of the viewing angle $\theta\in$ [0\textdegree,90\textdegree] is divided into 30 bins. Afterward, the distribution of $N_H$ is computed by randomly selecting $N_H$ 10,000 times from the sets of 192 models and 30 bins of $\theta$. The top panel of Fig.~\ref{fig:stat} shows the histogram of the $N_H$ fraction in a given bin. Based on the random selection, the fraction of Type 1, CN Type 2s, and CK are 0.56, 0.23, and 0.21, respectively. For obscured AGNs, the fraction of AGNs peaks near $\log[N_H(\percmsqr)]\approx 23.5$. The bottom panel of Fig.~\ref{fig:stat} represents the scattering of $\theta$ for obscured AGNs when they are observed randomly. The solid data points correspond to CN AGNs and the starred ones are for CK AGNs. The data points with $\theta\la 20$\textdegree\, near $N_H$=10\tsup{22.5} \percmsqr\, are gas dominated NSDs where starbursts did not occur. All the models appeared to be CK when they are viewed edge-on. This is expected since dense gas occupies the mid-plane regions of the discs. The figure illustrates that a given $N_H$ of CK or a heavily obscured CN AGN spans almost the entire domain of $\theta$ (from 0\textdegree\,to $\approx$80\textdegree) dependent on physical conditions. That means that, under certain conditions, the AGNs can be obscured by near face-on CK dusty gas.

The 2D NSD theory predicts $\fCN$ and $\fCK$ to be 23\% and 21\%, respectively, which is shown in the top panel of Fig.~\ref{fig:stat}. This is in reasonable agreement with the values mentioned in the literature. For instance, the modeling of CXB suggests that $\fCK$ can range from 5\% to 50\% \citep{akylas12}. \cite{brightman12} found $\fCK$ $\approx 0.25$ around redshift 1 by studying the sample of \textit{Chandra Deep Field South} survey. In the local universe, $\fCK$ is $\approx20\%$ based on the hard X-ray \citep{burlon11}, the optical \citep{akylas09}, and IR \citep{brightman11a,brightman11b} samples. \cite{ueda14} constraints the ratio of CK to CN AGNs to be $\approx0.5-1.6$ in order to produce the 20-50 keV part of the CXB spectrum which is in the agreement with the prediction of the NSD theory, $0.9$. The outcomes of our work is in good agreement with observations and suggests that NSDs could be potential sources of AGN obscuration at intermediate redshifts.

\subsection{Inversion Phenomena}
Many of our hydrostatic models of NSDs show a density inversion phenomenon. For example, the phenomenon is observed in model A whose vertical profiles at $\Rcric$ are shown in Fig.~\ref{fig:modelA_vertical}. There are two causes for the density inversion: (1) sublimation of grains and (2) demand of efficient energy flux transport. The latter case has been widely studied by theorists \citep{chitre67,joss73,schwarzschild75,tuchman78,harpaz84}. This inversion phenomenon can lead to a possible Rayleigh-Taylor instability (RTI). The development of RTI in radiation pressure dominated dusty environment has been also observed in hydrodynamical numerical simulations \citep[e.g.,][]{krumholz12,davis14}. We expect the density inversion due to the sublimation of grains to have an insignificant effect on $N_H$ distributions (thus also $\fCN$ and $\fCK$) since the inverted column density of gas is a small fraction (< 1\%) of the total column density.

The density inversion in gas pressure and turbulence pressure dominated layers causes a local pressure inversion. The development of a pressure inversion is expected in thermally unstable regions when temperature is monotonically increasing with $\tau$ \citep[e.g.,][]{rozanska96}. Such a development of inverted pressure has been already reported in the modeling of astrophysical atmospheres in the past \citep{achmad97,asplund98,helling01,rozanska02}. In order to analyze the effect of the inverted gas on a NSD, a time-dependent calculation becomes necessary which is beyond the scope of this paper. However, our comprehensive conclusion that NSDs are plausible sources for AGN obscuration is still absolute since the loss of dust opacity can cause an inflation to a parsec/sub-parsec atmosphere by orders of magnitude which is shown in the panel (b) of Fig.~\ref{fig:modelA_vertical}.

\subsection{Vertical Distribution of $\Sigmastardot$}
\label{sec:verticaldistribution}
Star-formation (SF) does not necessarily have to occur only at the mid-plane. Therefore, it would be interesting to study how the vertical distribution of SF affects the scale-height $h$ and $N_H$ distribution of NSDs. The Kennicutt-Schmidt law (which is also called the star-formation law) relates star-formation rate density $\Sigmastardot$ to the surface gas density $\Sigma_{\text{gas}}$ by $ \Sigma_{\text{gas}} \propto \dot{\Sigma}^\alpha_*$ \citep{schmidt59,kennicutt98,kennicutt12}, where $\alpha$ is expected to be $\approx 1$ for radiation pressure dominated regions \citep{ballantyne13}. Hence, we model the linear vertical distribution of $\Sigmastardot$ on $\Sigma$ using the prescription presented by \cite{hubeny98}. Implementation of the vertical distribution assumes: (1) a surface scale-height at $\tau=1$ and (2) $\tau_{\text{mp}}$>1.0 (For optically thin column of gas, SF is only allowed to occur at the mid-plane.).
With this description, we present the result for model A with the vertical distribution of $\Sigmastardot$ (hereafter, model $\barA$).

\begin{figure}
\includegraphics[width=0.46\textwidth]{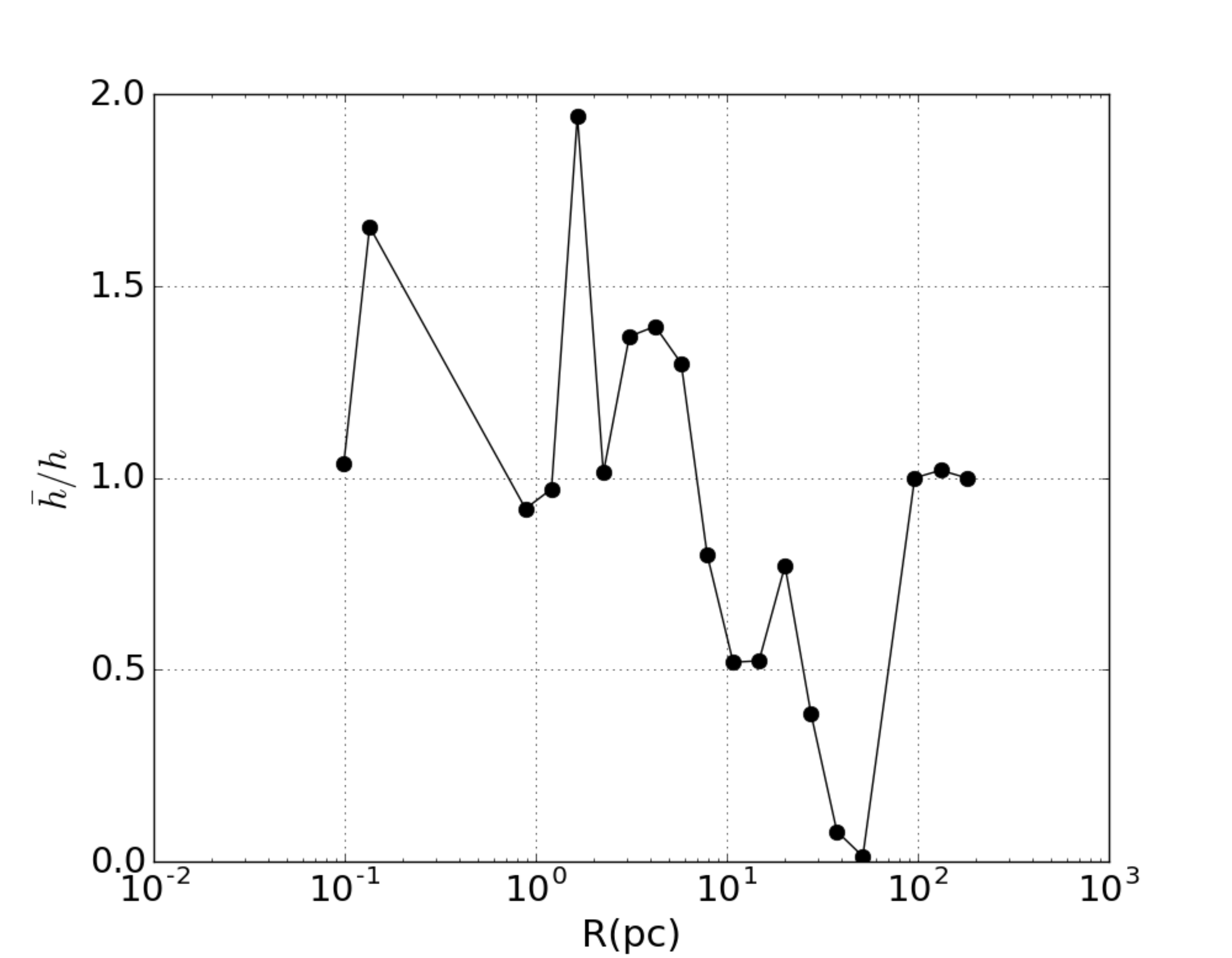}
\centering
\caption{Illustrating the effect of vertical distribution of $\Sigmastardot$ on the surface scale-height. $\bar{h}$ (model $\bar{\text{A}}$) has a slight variation in comparison to $h$ (model A) which are due to the following reasons: (1) a distribution of SF (model $\barA$) decreases density near the mid-plane for radiation pressure dominated region in compare to model A, (2) $\Pturb$ is approximated with a constant $\vturb$ when SF is only allowed to occur at the mid-plane, and (3) a difference in the amount of pressure inversion affects the difference in the scale-heights computed from both models.}
\label{fig:height_vd}
\end{figure}

Fig.~\ref{fig:modelA_radial} (dashed-dotted-green line) shows the radial distribution of the surface scale-height $\bar{h}$ (for model $\barA$) and Fig.~\ref{fig:height_vd} plots the ratio of $\bar{h}$ to $h$ (for model A) as a function of $R$. For $R\leq 5$ pc, the atmospheres are entirely dominated by radiation pressure. On sub-parsec scale ($<$1.5 pc), grains are sublimated and the opacity gradient induced due to the dust sublimation mainly governs the surface scale-height. Therefore, $h$ is approximately $\bar{h}$ in this region. For $1.5<R<5$ pc, the region is still radiation dominated, but the maximum temperature is below the sublimation temperature of dust. Here, model $\barA$ possesses a lower density region than model A which results in a slightly higher $\bar{h}$ than $h$. For $R>5$ pc, the region is mostly turbulence pressure dominated for model A and gas pressure dominated for model $\barA$. In these regions, the surface scale-height is primarily set by the pressure inversion phenomena. Model $\barA$ introduces higher pressure inversion than model A. Increasing pressure inversion causes increase in density \citep{helling01} which results in decreasing $\bar{h}$ in comparison to $h$. At the outer radii, (last three slabs), a column of gas is optically thin; therefore, $\bar{h}/h=1.0$. Another source of a difference in $\bar{h}$ and $h$ is a difference in treatment of turbulence pressure in both models. For model A, $\Pturb$ was approximated by assuming a constant $\vturb$, while, for model $\barA$, $\Pturb$ was computed exactly using Eq.~\ref{eqn:pturb}. In summary, the vertical distribution of SF has a small effect on the surface scale-height. We also checked the $N_H$ distributions from models A and $\bar{A}$ and their difference is negligible. In general, one expect to have a minimum effect of $\Sigmastardot(\Sigma)\propto\Sigma$ on the vertical structure. That is because a distribution of energy source only affects the Hopf function $q(\tau)$ which is introduced in the temperature solution of RTE and $T(\tau)$ has a weak dependance on the linear distribution of $\Sigmastardot$.
\subsection{Caveats \& Directions for Future Work}
One of the important tasks in modeling a physical system is to be aware of possible caveats which can be introduced due to assumptions, approximations, or an exclusion of relevant mechanisms. In the presented framework of a NSD, due to the assumptions of diffusion approximation and grey problem, results from modeling may be inaccurate for an optically thin limit ($\tau<<1$). However, this does not affect our conclusion that NSDs posses an inflationary atmosphere on parsec/sub-parsec scale which can potentially obscure the AGN irradiation since $h\sim R$ is achieved before $\tau$ reaches the optically thin limit. Another implicit assumption due to statistics based on a random selection is that there is not any internal/natural biased in the sample of input parameters. However, this assumption may be imprecise; for instance, the black hole mass distribution function is not evenly weighted for a given redshift. Another caveat in the modeling is that it misses outflows. In star-forming regions, outflows are possible to occur due to stellar feedback \citep{leitherer92}, supernovae feedback \citep{chevalier85}, and radiation pressure on dusty atmosphere \citep{murray05}. Moreover, outflows are observed in many star-forming galaxies with the speed of $>1000$km s\tsup{-1} at $z\sim 0.6$ \citep{tremonti07,diamond12}. In the sample of local starbursts, outflows with the speed of 1500 km s\tsup{-1} are observed by \cite{heckman11}. In the future, we may study our sampled models using Bayesian statistics in order to remove an intrinsic biased and compare its conclusion with statistics based on a random drawings.

Another missing ingredient in the modeling is the effect of AGN (if they are present in galaxies) on the NSD structure. To study its importance on the presented modeling, we compute the optical depth along line of sights ($\tau_{\text{los}}$) for $\theta$ above the dust sublimation layer at the critical radius and check if $\tau_{\text{los}}$ is greater than the unity. Only 16 out of 99 models are optically thin and being transparent to the ultraviolet (UV) radiation. Depending on the physical conditions of the NSD, the UV energy can be deposited near the surface or in the deeper region along a line of sight since $\tau_{\text{los}}$ can be as large as 10$^4$. Moreover, we can compute the dust sublimation radius ($r_{\text{s}}$) with the following equation \citep{barvainis87}:
\begin{align}
r_s=1.3\Bigg(\frac{L_{\text{UV}}}{10^{46}\text{erg s\textsuperscript{-1}}}\Bigg)^{1/2}\Bigg(\frac{T_s}{1500 \text{K}}\Bigg)^{-2.8} \text{pc}.
\end{align}
Here, $L_{\text{UV}}$ is the UV luminosity of the AGN and $T_s$ is the sublimation temperature. Using the sublimation temperature of graphite grains (1750 K) and a typical bolometric luminosity (which gives the upper limit on $r_s$) range of Seyfert galaxies (10$^{42}$ to 10$^{44}$ ergs s$^{-1}$), the approximated $r_s$ ranges from 0.008 to 0.08 pc, whereas Fig.~\ref{fig:height} shows that the critical radius (where the largest expansion occurs in a disc) ranges from 0.1 to 2.3 pc. This implies that the AGN heating does not completely destroy the grains which are residing at the surface of the NSD at parsec scale. Based on this rough estimate, the AGN heating may not have a significant effect on the presented hydrostatic structure. However, a proper treatment of the AGN irradiation is required to be conclusive which will be done in detail in future work. Finally, NSDs may be present in galaxies with no central AGNs; therefore, this model may have wider applicability beyond AGNs.

Based on \textit{Hubble Space Telescope} and ground based telescopes, nuclear star clusters (NSCs) are observed to be a common phenomena occurring in many types of galaxies: 70\% in spheroidal galaxies \citep{cote06} and 75\% in late-type (Sc-Sd) disc galaxies \citep{boker02,walcher05}. Their sizes are at the parsec scale, 2-5 pc \citep{geha02,boker04,cote06}. Two possible scenarios are proposed for the formation of NSCs: (1) merger of dense clusters via dynamical friction \citep{andersen08,capuzzo08} and (2) gas accretion at the parsec-scale \citep{mastropietro05,seth06}. NSDs fall under the second scenario which can be a potential progenitor of NSCs. The NSD theory predicts a compact star-forming region with intense SF rate ($>3M_{\astrosun}$year\tsup{-1}) on parsec scale. Moreover, the 2D calculation of NSDs show these regions possess an environment whose scale-height is approximately larger by an order of magnitude than the radial distance. These characteristics of NSDs pose an interesting question: can nuclear star clusters be remnants of nuclear starburst discs? We will be addressing this question in the future investigation.

\section{Conclusion}
To summarize, we have successfully developed an iterative algorithm to compute the 2D hydrostatic structure of a nuclear starburst disc. These discs can be a potential source in obscuring the AGN radiation when an abundant amount of gas is available in galaxies at intermediate redshift fueling the central black hole. This modeling allows us to confirm that NSDs can possess the radiation pressure supported dusty structure on sub-parsec/parsec scale covering a major part of the sky (observed from a central back hole). Our results show that the NSD with fixed physical conditions can appear to be obscured by CN or CK gas depending on a viewing angle which supports the basic unification theory of AGNs. Below we summarize our main findings:
\begin{itemize}
\item A starburst phenomenon is more likely to occur in a disc with a smaller size, larger gas fraction, and higher Mach number.
\item 52\% (99/192) of models show the large expansion of an atmosphere on parsec/sub-parsec scales. This indicates that a large expansion at the parsec/sub-parsec scale is a common phenomenon in dusty star-forming regions.
\item Within the input parameter space, the atmosphere can be expanded from 0.2 to $\sim$ 30 $h/R$ at the critical radius ($\Rcric$) ranging from $\sim$0.2 to $\sim$2 pc. For a lower black hole mass, the inflation occurs closer to the black hole as indicated in Fig.~\ref{fig:height}.
\item Based on the random sample of input parameters, the NSD theory predicts 56\% of Type 1, 23\% of CN Type 2s, and 21\% of CK AGNs. Within the sample of obscured AGNs, the distribution of CN Type 2s peaks near 10\tsup{23.5} cm\tsup{-2}. These predictions are consistent with observational evidence \citep[e.g.,][]{akylas09,burlon11,brightman11a,ueda14}.
\item Based on the 2D NSD theory, $N_H$ along a line of sight varies from 10\tsup{22.5} \percmsqr\, to 10\tsup{29.5} \percmsqr. From an edge-on view, all NSD models appear to be CK which is expected since dense gas resides at the mid-plane.
\item Our results show a heavily obscured CN and CK AGNs span almost the entire range of $\theta$ from 0\textdegree\, to $\approx$80\textdegree. The given AGN-NSD system (fixed input parameters) appears to be CN to CK for $\theta$ less than $\theta_{\text{max}}$; otherwise, the central AGN would appear to be Type 1. This supports the basic unification theory of AGNs. The restrictions on $N_{H,\text{max}}$ and $\theta_{\text{max}}$ of this system is determined by its physical conditions (input parameters).
\item A linear vertical distribution of $\Sigmastardot$ has an insignificant effect on $N_H$ distributions since the distribution of SF only affects the temperature solution of RTE (Eq.~\ref{eqn:tempsolution}) and its effect is negligible.
\end{itemize}

\section*{Acknowledgement}
This work is supported by NSF award AST 1333360. The authors are thankful to the referee for useful comments.

\newpage
\bibliographystyle{mnras}
\bibliography{refs/ref_a,refs/ref_b,refs/ref_c,refs/ref_d,refs/ref_e,refs/ref_f,refs/ref_g,refs/ref_h,refs/ref_i,refs/ref_j,refs/ref_k,refs/ref_l,refs/ref_m,refs/ref_n,refs/ref_o,refs/ref_p,refs/ref_q,refs/ref_r,refs/ref_s,refs/ref_t,refs/ref_u,refs/ref_v,refs/ref_w,refs/ref_x,refs/ref_y,refs/ref_z,refs/ref_books}{}
\bibliographystyle{mnras}

\appendix
%\section{Appendicies}
\section{Rosseland Mean Opacity}
\label{appendix:rmo}
Modeling of radiation pressure dominated regions, such as NSDs, requires one to take into account the interaction between the radiation field and matter. This interaction can be quantified with a macroscopic quantity, opacity. A computation of opacity is not a trivial task since each chemical element interacts differently with a different frequency of light. Moreover, the inclusion of dust in the astrophysical composition increases the complexities due to the variability (e.g., size, shape, and distribution of grains) of dust opacity. However, for an optically thick medium (optical depth $\tau>>1$), the opacity of a medium can be well treated as the Rosseland mean opacity ($\kappa_R$) \citep{book_hubeny14} where the radiation pressure can be well described by the diffusion approximation \citep{book_hubeny14}. Under the condition of LTE, $\kappa_R$ depends on the composition of matter, its temperature $T$, and the number density $n_H$. For NSDs, we choose Orion abundances to compute opacity under LTE using CLOUDY \citep{ferland13}. CLOUDY is a one dimensional photo-ionization code which includes grain physics \citep{weingartner01,vanhoof04,weingartner06} with a realistic size distribution of various grain species \citep[e.g.][]{mathis77}; for instance, the size and shape based absorption and scattering processes of grains. For NSDs, we explore a wide range of temperature $T$ and density $n_H$:
\begin{center}
\begin{align}
& 10\text{K}\leq T \leq 10^{6.1}\text{K}\\
& 10^{-7}\text{\percmcube} \leq n_H \leq 10^{13}\text{\percmcube}
\end{align}
\end{center}

\begin{table*}
\begin{center}
    \begin{tabular}{ | l | l | l |l|l|p{2cm} |}
    \hline
    Region & I & A & B & C & D \\ \hline
    $\log(n_H)$ & Independ. & [3.0,6.0] & [6.0,13.0] & [3.0,13.0] & [-7.0,3.0]\\ \hline
    $\Delta$ $\log(n_H)$ & N/A & 0.25 & 0.5 & 0.5 & 0.25 \\ \hline
    $\log(T)$ & [1.0,3.0] & [3.0,4.1] & [3.0,4.1] &[4.1,6.1]& [3.0,6.1] \\ \hline
    \end{tabular}
    \caption{2D interpolation of $n_H$ and $T$. The resolution among temperature points is taken to be 0.01 except
    for ranges of [3.13,3.26] and [3.243,3.2431] which has resolutions of 0.005 and 0.0001, respectively.}
    \label{tab:cloudydomain}
\end{center}
\end{table*}

\begin{figure}
\includegraphics[width=0.55\textwidth]{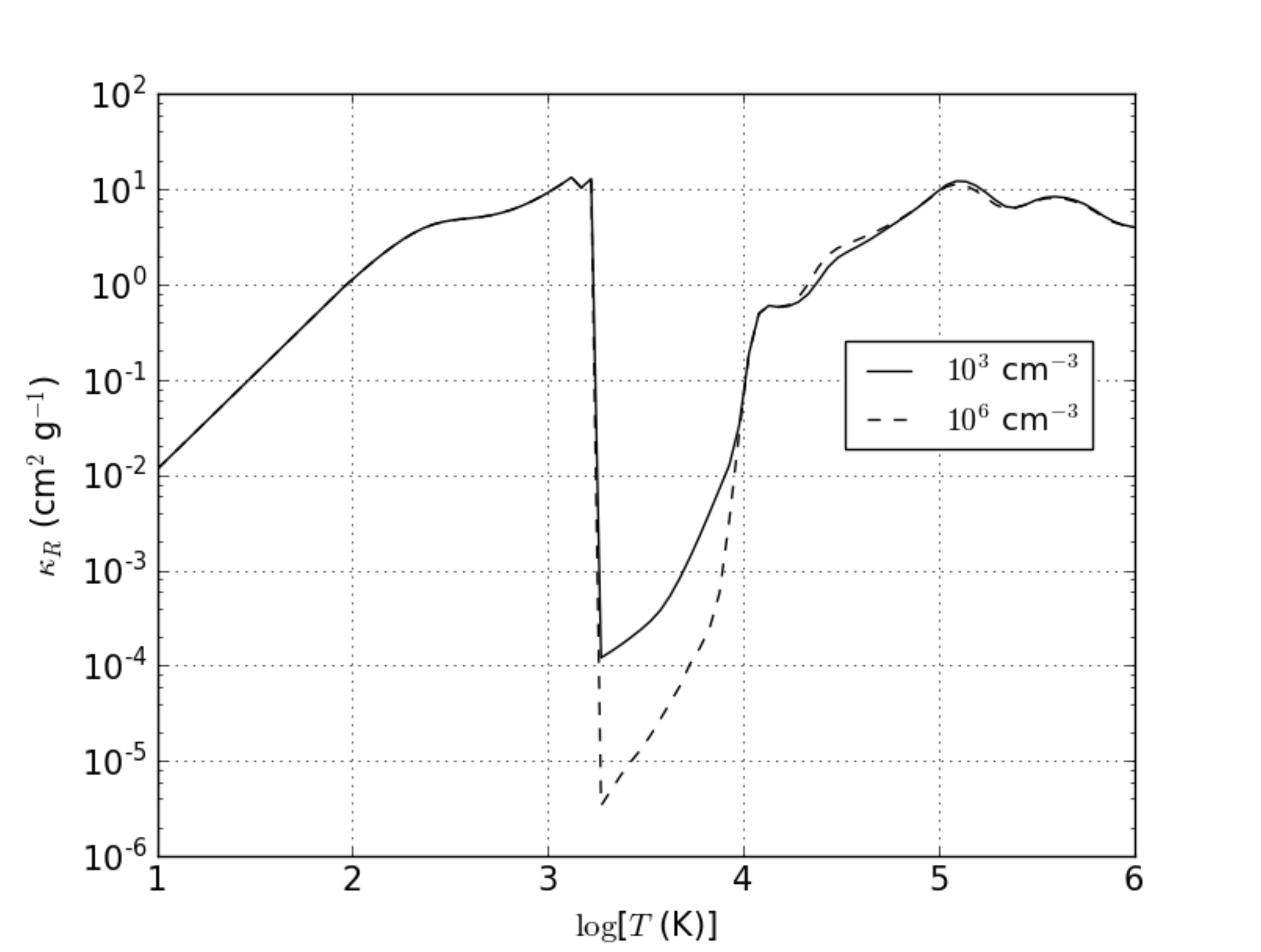}
\caption{Rosseland mean opacity ($\kappa_R$) as a function of temperature. Solid and dashed lines represent for $\log[n_H($\percmcube$)]=$ 3 and 6, respectively. In low-temperature region, the opacity is dominated by grains which drops near 1750K when graphite grains are destroyed. Due to the decrease in ionization factor with density, an overall opacity decreases as a function of density for a fixed temperature near $\log[T(\text{K})]=3.5$. In high temperature limit, $\kappa_R$ is dominated by the Thompson scattering, thus it becomes nearly independent of density.}
\label{fig:rmo}
\end{figure}

The curves in Fig.~\ref{fig:rmo} show $\kappa_R$ as a function of $T$ for $n_H$ of 10\tsup{3} cm\tsup{-3} and 10\tsup{6} cm\tsup{-3}. In low temperature regime, the $\kappa_R$  is increasing with the temperature for a given $n_H$. This is
expected since dust is the dominant
source of opacity in this region and the dust cross-sectional area is proportional to the wavelength $\lambda$ of incoming electromagnetic wave by $\lambda^{-\alpha}$, where $\alpha$ approaches to 2 in the Rayleigh limit \citep{pollack85}. (As temperature increases, the peak of radiation power shifts toward a lower $\lambda$).
The $\kappa_R$  has a small drop near 1400K \citep{kishimoto13} due to the sublimation of silicate grains and the second deep drop occurs near 1750K \citep{kishimoto13} due to the sublimation of graphite grains.
The increase in $\kappa_R$ after the deep drop is due to the H- scattering opacity then subsequently bound-free and
free-free opacity. Hydrogen starts ionizing near 5000K, and further ionized by $\sim 80\%$ and $\sim 100\%$ near 10\textsuperscript{4}K and 1.5$\times$10\textsuperscript{4}K, respectively. Using Saha's equation, one can show an ionization fraction decreases (higher bound-free opacity) with density. This is responsible for an overall decreased in $\kappa_R$ with $n_H$ in the intermediate temperature regime since hydrogen is the abundant element in the composition. After hydrogen is fully ionized, the opacity is fairly independent of density since Thompson scattering becomes dominant.
As temperature increases, gas absorbs more energy from incoming radiation which increases overall opacity of the medium.

Based on the complexities and behaviors of the Rosseland mean opacity as shown in Fig.~\ref{fig:rmo}, the two dimensional space of density and temperature is divided into five regions to have an optimal algorithm for interpolation. The logarithmic grids and its resolution for each region are described in Table~\ref{tab:cloudydomain} and were
chosen such that it optimizes the computational time and accuracy of the interpolation algorithm. We choose a linear interpolation among logarithmic grid points. For region I, $\kappa_R$  is an independent of density as seen in Fig.~\ref{fig:rmo}. Therefore, the $\kappa_R$  of $\log[n_H (\percmcube)]$ = 3.0 is taken for whole density range in interpolation function and only temperature grid points are interpolated.
\begin{figure}
\includegraphics[width=7.5cm]{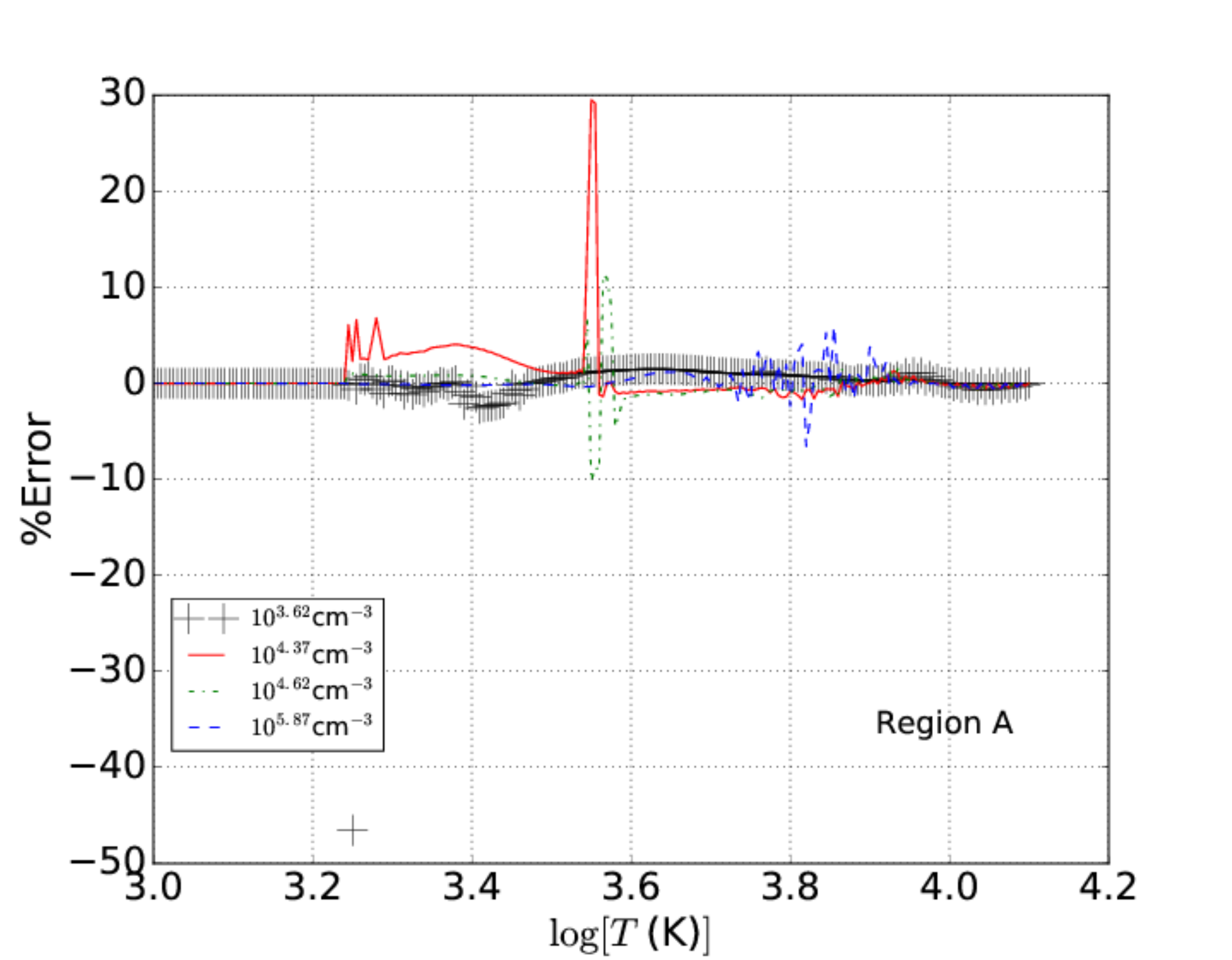}
\includegraphics[width=7.5cm]{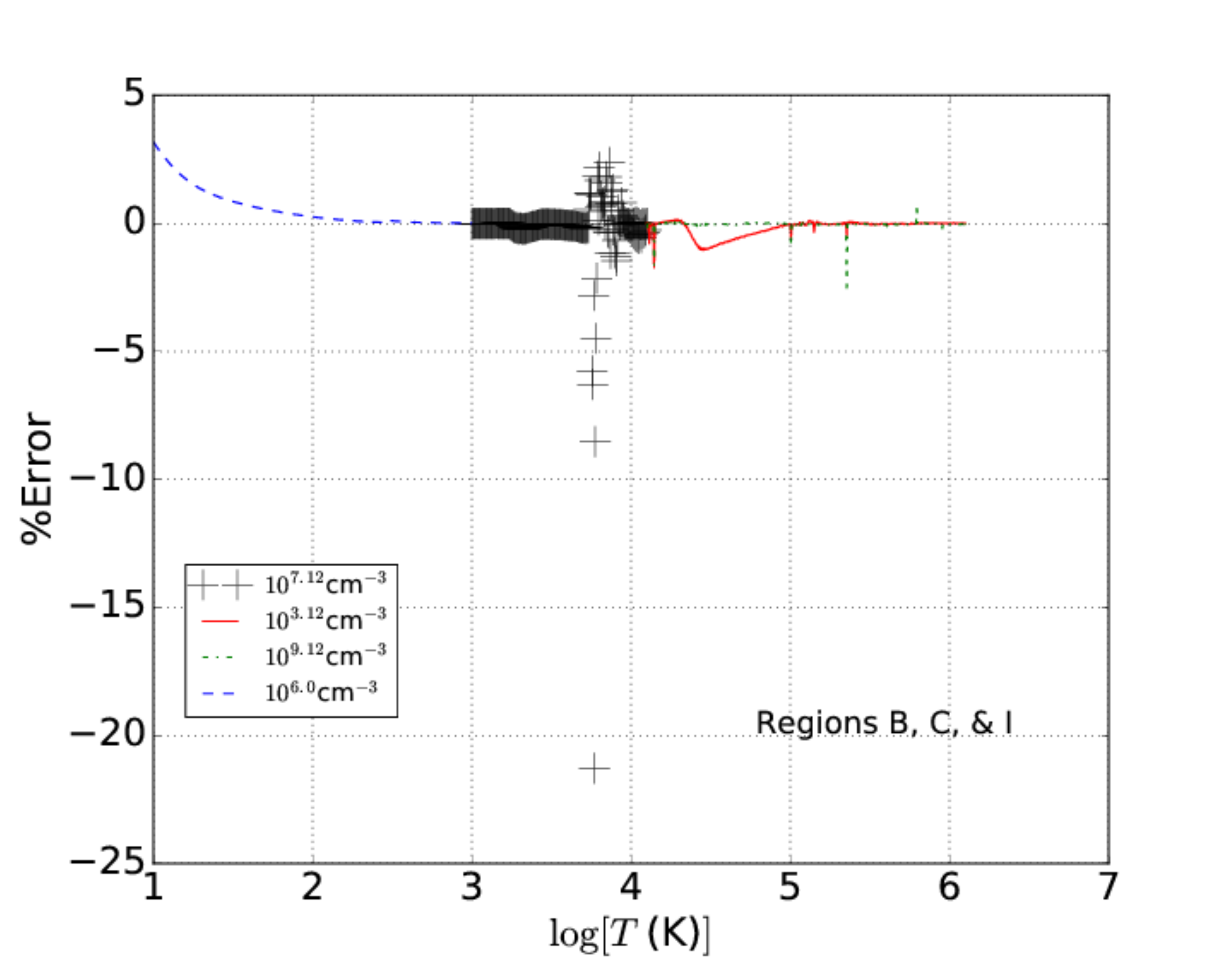}
\caption{Representing the analysis on the accuracy of interpolation algorithm. For regions I and C, uncertainty in predicted opacity
is less than 4\%. For regions A and B, error is less than $\sim$ 10\% except for certain $(\log(n_H),\log(T))$ sets, yet less than 50\%.}
\label{fig:cloudy_interpolation}
\end{figure}

To check the accuracy of the interpolation algorithm, opacities from the algorithm were predicted for various temperature and
density (off the grids) points and the discrepancy against the real data from CLOUDY was computed as
\begin{center}
\begin{align}
\text{\%Error}\equiv\frac{100\times (\text{Predicted value-Actual value)}}{\text{Actual value}}.
\end{align}
\end{center}
Examples of the error analysis is illustrated in Fig.~\ref{fig:cloudy_interpolation}. The percentage error in regions I and C are less than 4\%. For regions A and B, the spacing of grid points is not linear because of the deep drop in $\kappa_R$  due to the sublimation of graphite grains. To minimize the percentage error in region B, various resolutions are chosen for subspaces. In region A, the interpolation function gives the percentage error that is mostly less than 10\% except near $\log[T(\text{K})]\sim$ 3.3 and 3.6, where the error reaches 50\% and 30\%, respectively. Region B shows less than 4\% error except it can reach up to 25\% near 10\tsup{3.8}K. In summary, every vertical slab may have few layers with an inaccurate opacity (yet less than 50\% uncertainty). However, this will not have a significant effect on the overall structure for the purpose of this project since the number of these layers ($\approx$ 2-5) is a small fraction of total grid points (300+).

%%%%%%%%%%%%%%%%%%%%%%%%%%%%%%%%%%%%

\bsp % ``This paper has been produced using the ...''

\label{lastpage}
\end{document}